\UseRawInputEncoding 

\documentclass{nature-pre}
\usepackage{amsmath}
\usepackage{amssymb}
\usepackage{graphicx}

\usepackage{notes2bib}

\usepackage{underscore}
\usepackage[english]{babel}

\usepackage{multirow}

\usepackage{tabularx} 

\usepackage{comment}

\graphicspath{./}
 


\title{Dipolar spin relaxation of divacancy qubits in silicon carbide}

\author{Oscar Bulancea Lindvall$^{1}$, Nguyen T. Son$^{1}$, Igor A. Abrikosov$^{1}$, Viktor Iv\'{a}dy$^{1,2,3,*}$}

\begin{document}
\maketitle

\begin{affiliations}
\item   {Department of Physics, Chemistry and Biology, Link\"oping
  University, SE-581 83 Link\"oping, Sweden}
\item  {Wigner Research Centre for Physics, Hungarian Academy of Sciences,
  PO Box 49, H-1525, Budapest, Hungary}
 \item   {Max-Planck-Institut f\"{u}r Physik komplexer Systeme, N\"{o}thnitzer Street 38, D-01187 Dresden, Germany}
\item[*] Correspondence should be sent to viktor.ivady@liu.se
\end{affiliations}

\date{\today}

\newpage

\begin{abstract}
Divacancy spins in silicon carbide  implement qubits with outstanding characteristics and capabilities in an industrial semiconductor host. On the other hand, there are still numerous open questions about the physics of divacancy point defects, for instance,  spin relaxation has not been thoroughly studied yet. Here, we carry out a theoretical study on environmental spin induced spin relaxation processes of  divacancy qubits in 4H-SiC. We reveal all the relevant magnetic field values where the longitudinal spin relaxation time T$_1$ drops resonantly due to the coupling to either nuclear spins or electron spins. We quantitatively analyze the dependence of the T$_1$ time on the concentration of point defect spins and the applied magnetic field in the most relevant cases and provide an analytical expression. We demonstrate that dipolar spin relaxation plays a significant role  both in as-grown and ion implanted samples and it often limits the coherence time in 4H-SiC. 

\end{abstract}

\newpage


\section*{Introduction}

Through the example of the nitrogen-vacancy center in diamond\cite{duPreez:1965,Wrachtrup:JPCM2006,Maze2011,DohertyNVreview} (NV center), point defects in wide-band gap semiconductors have demonstrated their potential for quantum-enhanced technologies. In particular, the NV center based devices are about to revolutionize sensing at the nanometer scale\cite{
Balasubramanian:Nature2008,Taylor:NatPhys2008,Kucsko2013,PlakhotnikNanoLett2014,Schirhagl2014,DegenRMP2017,glenn_high-resolution_2018,schmitt_submillihertz_2017}, while NV centers coupled to adjacent nuclear spins can serve as nodes for quantum internet\cite{bernien_heralded_2013,pfaff_unconditional_2014,wehner_quantum_2018} and quantum computations\cite{Weber10,Kurizki2015,ZhangPRL2020}. While the physics of the NV center is understood to a very large degree, other point defect qubits with similar or even superior capabilities, such as the neutral silicon and carbon vacancy pair (divacancy)\cite{Koehl11,Christle2014,Son2006} and the negatively charged silicon vacancy (silicon vacancy)\cite{Soltamov12,Widmann2014} in silicon carbide (SiC) are still subjects of active fundamental studies. For example, recent reports on the neutral divacancy have demonstrated  64 ms coherence time\cite{miao_universal_2020}, implementation of spin-to-photon interface\cite{ChristlePRX2017}, nuclear spin operations for quantum memory applications\cite{FalkPRL2015,bourassa_entanglement_2020} and dynamic nuclear polarization\cite{FalkPRL2015,IvadyDNP2015,IvadyPRL2016},  and a room temperature spin contrast as high as 30\%\cite{li_room_2020}. The fact that SiC is a technologically mature semiconductor with controllable p- and n-type doping and  established nanofabrication techniques further enhances its importance and paves the way toward affordable integrated quantum devices\cite{whiteley_spinphonon_2019,anderson_electrical_2019,son_developing_2020}.

Besides coherent properties of the qubits, longitudinal spin relaxation, with the corresponding decay time T$_1$, is of great importance as it sets the fundamental limit for several applications, e.g. for dynamic decoupling techniques and sensing.\cite{pham_enhanced_2012,bar-gill_solid-state_2013,RomachPRL2015,barry_sensitivity_2020} Relaxation of the NV center's electron spin at different magnetic fields, temperatures, and environmental spin defect concentration has been studied.\cite{takahashi_quenching_2008,jarmola_temperature_2012} Furthermore, optical signatures of strong environmental couplings were revealed, enabling novel microwave free spectroscopy  applications.\cite{Wang2014,WickenbrockAPL2016,WoodPRB2016,WoodNatComm2017,ivady_photoluminescence_2021} In contrast, much less is known about the longitudinal spin relaxation of divacancy qubits in SiC.  The T$_1$ time was reported for specific values of the magnetic field at various temperatures \cite{koehl_room_2011,falk_polytype_2013,yan_room-temperature_2020}, however, no systematic studies on the environmental couplings and resulting spin relaxation processes have been carried out yet. Even less is known about environmental resonances that may give rise to T$_1$ spectroscopy and dynamic nuclear polarization in SiC.

In this article, we study the longitudinal spin relaxation of divacancy qubits in 4H-SiC due to various environmental spins, such as spin-1/2, spin-1, and spin-3/2 point defects and $^{13}$C and $^{29}$Si nuclear spins. First, we identify the most relevant resonance magnetic field values where efficient coupling to the environmental spins leads to enhanced spin relaxation, and second, we quantitatively study the spin relaxation rate, T$_1^{-1}$, over a wide range of magnetic field values, with two orders of magnitude variation of the defect concentration, and various isotope abundances. We provide a simple analytical formula with theoretically fitted constants that can be used to obtain ensemble averaged T$_1$ time in a sample of given defect concentration and isotope abundance.

\section*{Background}
\label{sec:back}

\subsection{Theory of longitudinal spin relaxation}
\label{sec:relax}

For a given quantisation axis,  specified for example by the direction of the external magnetic field,  the magnetization of an integer or half-integer spin can be described by the longitudinal and the transversal components.  The longitudinal,  field parallel,  magnetization is time independent for isolated spins, while the transversal, perpendicular to magnetic field,  magnetization component rotates around the quantization axis with the Larmor frequency $\omega = -\gamma B$,  where $\gamma$ is the gyromagnetic ratio and $B$ is the external magnetic field.  

Due to the interaction of the spin with its environment,  the above dynamics is altered in two ways,  the longitudinal magnetization becomes time dependent and the Larmor frequency may be perturbed,  i.e.\ the spin accumulates a phase shift compared to the isolated case during its precession around the magnetic field.  While the total magnetization of a single spin is always preserved, ensemble measurement and time averaged quantities exhibit the  relaxation of both the longitudinal and transversal magnetization due to the interaction with the environmental.  For example, the longitudinal magnetization of an ensemble vanishes with time and approaches the thermal state with zero magnetization, where the spins are pointing in any direction with equal probability.  In a large enough ensemble, the longitudinal magnetization relaxes exponentially to the thermal state with relaxation time T$_1$.  Similarly, the accumulation of varying phase shifts either in an ensemble or over time gives rise to the reduction of the net transversal magnetization,  often referred to as dephasing or decoherence.

 In order to quantitatively describe spin relaxation and decoherence of spin a ensemble in an open quantum system, one may use the density operator that can be expressed as
 \begin{equation}
 \hat{\varrho} = \sum_n^N \left| \phi_n \right\rangle  \! \!  \left\langle \phi_n \right| \text{,}
 \end{equation}
where $\phi_n $ the wavefunction of spin $n$ in an ensemble of $N$ spins.  The density matrix of spin-1/2 objects can be written in a given quantization basis as a $2 \times 2$ matrix
\begin{equation}
\varrho = 
\begin{pmatrix}
\varrho_{\uparrow \uparrow} & \varrho_{\uparrow  \downarrow}   \\
\varrho_{ \downarrow \uparrow }  &  \varrho_{\downarrow \downarrow}
\end{pmatrix} \text{,}
\end{equation}
where $\varrho_{ij} = \left\langle  i  \right|    \hat{\varrho} \left| j\right\rangle = \sum_n   \left\langle i  \middle|  \phi_n \right\rangle  \! \!  \left\langle \phi_n \middle|  j \right\rangle$, where $\left| i\right\rangle$ and $\left| j\right\rangle$ are the eigenstates of the spin $z$ operator $S_{z}$ and $i,j = \left\lbrace \uparrow, \downarrow \right\rbrace$.  The diagonal elements of the matrix  specify the net longitudinal magnetization of the spin ensemble as $\bar{\mu}_{\parallel} = g_e \mu_B \left( \varrho_{\uparrow \uparrow} - \varrho_{\downarrow \downarrow } \right)/2$, while the  off-diagonal elements of the density matrix specifies the coherence properties of the ensemble \cite{yang_quantum_2008,seo_quantum_2016,onizhuk_probing_2021}.  In case of a spin-1 ensemble the density matrix can be expressed by a $3\times 3$ matrix whose diagonal elements $\varrho_{k} = \varrho_{kk} $, where $k = \left\lbrace -1,0,+1 \right\rbrace$, define the longitudinal magnetization.

In this paper we focus only on the longitudinal spin relaxation of  point defect spins.  One may find further information on the theory and calculation of the decoherence in Refs.~[\cite{yang_quantum_2008,seo_quantum_2016,onizhuk_probing_2021}]. In semiconductors, there are two main contributions to the longitudinal spin relaxation dynamics of point defect qubits.  Spin-orbit coupling and phonon scattering assisted processes are dominating the temperature dependence and are the most efficient relaxation processes at high temperature in general.  Due to this reason the phenomena of  longitudinal spin relaxation is frequently referred to as spin-lattice relaxation in the magnetic resonance community.  

In general,  there are several contributions to the spin relaxation rate\cite{shrivastava_theory_1983,norambuena_spin-lattice_2018} ($1/T_1$) 
that can be written as
\begin{equation}
\frac{1}{T_1} = \frac{1}{T_{1}^{dd}}  + A_{\text{dir,0}} + A_{\text{dir,1}} \mathcal{T} + A_{\text{Ram}}\mathcal{T}^n  + \frac{A_{\text{Orb}}}{e^{\Delta/k \mathcal{T}} -1 } + A_{\text{loc}} \frac{e^{\Delta_{\text{loc}}/k\mathcal{T}}}{\left( e^{\Delta_{\text{loc}}/k\mathcal{T}} -1 \right)^2} + A_{\text{therm}} \! \left( \mathcal{T}\right) \text{,}
\label{eq:rate}
\end{equation}
where $ 1/T_{1}^{dd}$ is the temperature independent spin-spin or dipolar relaxation rate,  the main subject of the  paper,  $A_{\text{dir,0}}$ and $A_{\text{dir,1}}$ are the coefficient of the temperature independent and the temperature dependent direct one phonon relaxation process, $\mathcal{T}$ is the temperature,  $A_{\text{Ram}}$ is the coefficient of the Raman process, $n = \left\lbrace 5, 6,7\right\rbrace$ is the exponent of the temperature dependence of the Raman process\cite{shrivastava_theory_1983,radczyk_applications_2003,hoffmann_raman_2013,norambuena_spin-lattice_2018},  $A_{\text{Orb}}$ is the coefficient of the Orbach  process,  $\Delta $ is the energy separation between the ground state and the excited state involved in the Orbach process,  $k$ is the Boltzmann constant,  $A_{\text{loc}}$ is the coefficient of the contribution from local-vibrational mode scattering,  $\Delta_{\text{loc}}$ is the energy of the local vibrational mode, and $A_{\text{therm}}$ describes thermally activated processes.  In wide band gap semiconductors, such as diamond and SiC the first four terms on the right hand side of Eq.~(\ref{eq:rate}) are the most relevant ones.\cite{redman_spin_1991,takahashi_quenching_2008,jarmola_temperature_2012,simin_locking_2017,soltamov_relaxation_2021,lin_temperature_2021}
Temperature dependence of the spin relaxation rate of divacancy qubits in SiC has 
been reported very recently.\cite{lin_temperature_2021} We fit Eq.~(\ref{eq:rate}) to the  data in Ref.~[\cite{lin_temperature_2021}] to quantify the strength of the various relaxation processes.  The fitted parameters for the dipolar spin relaxation, direct process, and Raman process can be found in Table~\ref{tab:parA}.  Orbach,  local, and temperature activated relaxation processes do not play a significant role in the reported temperature range for divacancy qubits.   We note that the temperature dependence in the 200-300~K interval is best fitted by $\mathcal{T}^7$ Raman processes\cite{norambuena_spin-lattice_2018}. Indeed, when $n$ is also varied in the least square fit, we obtain 6.953 for the exponent in the sample implanted with 10$^{14}$ cm$^{-2}$ N$_2$ ion dose.  Note also that the temperature independent spin relaxation contribution plays crutial role, and its value depends only slightly on the implantation dose.  From the fit we cannot determine the weights of the dipolar spin relaxation and the temperature independent direct one phonon relaxation processes.  The latter contribution is however sample independent.   Further, discussion on this observation  in the light of our results can be found in the Discussion section.

\begin{table}[!h]
\caption{Parameters of Eq.~(\ref{eq:rate}) for divacancy spin in 4H-SiC implanted with different dose of nitrogen molecule ion in Ref.~[\cite{lin_temperature_2021}]. $A_{\text{Orb}}$,  $A_{\text{loc}}$, and $A_{\text{therm}}$ are set to zero as the corresponding processes do not play a role. }
 \begin{center}
       \begin{tabular}{c|cccc} 
      N$_2$ ion dose (cm$^{-2}$)  & $\left( \frac{1}{T_{1}^{dd}}  + A_{\text{dir,0}} \right)$  (ms$^{-1}$) & $A_{\text{dir}}$ (ms$^{-1}$K$^{-1}$)  & $A_{\text{Ram}}$ (ms$^{-1}$K$^{-7}$)  & $n$ \\ \hline
      10$^{14}$ & 1.765 & 0.005866 & 2.715$\times 10^{-17}$  & 7\\
      10$^{13}$ & 1.575 & 0.004980 & 1.916$\times 10^{-17}$ & 7
       \end{tabular}
 \end{center}
\label{tab:parA}
\end{table}

Recent advances in first principles theory of spin-phonon relaxation processes have made parameter free calculations of temperature dependent spin relaxation rate possible,  see Refs.~[\cite{restrepo_full_2012,gugler_ab_2018,astner_solid-state_2018,park_spin-phonon_2020,xu_spin-phonon_2020}] for further details. The first principles simulation of dipolar spin relaxation of point defect spins has also recently been addressed in Ref.~[\cite{IvadyPRb2020}].  This relaxation process may be critical for shallow defects in nanometer scale sensing applications and for integrated devices, where the dipolar spin relaxation may enhance due to the increased doping level. Coupling to defects with various local spin Hamiltonians gives rise to a rich phenomena with numerous resonances where the spin relaxation rate can be enhanced by several orders of magnitude.\cite{jarmola_temperature_2012,IvadyPRb2020,ivady_photoluminescence_2021} In this paper we utilize this latter method to quantitatively study the temperature independent longitudinal spin relaxation rate of divacancy spin qubits in various spin environments.  In the following few paragraphs we shortly review the theory and basic features of this method.

Decoherence\cite{yang_quantum_2008,seo_quantum_2016} and dipolar spin relaxation are inherently many-body phenomena as they originate from the coupling of a central spin to its environment of countless other spins.  As the size of the Hilbert space exponentially increases with the number of environmental spins,  exact solution of a sufficiently large model is unfeasible.  Therefore,  approximations must be used.  Furthermore,  for point defect qubits the time scale of the longitudinal relaxation is often $10^6$-$10^8$ longer that the characteristic time scale of the coherent oscillations,  thus application of state-of-the-art quantum dynamics simulation approaches,  such as tDMRG, are impractical. 

\begin{figure}[h!]
\begin{center}
\includegraphics[width=0.90\columnwidth]{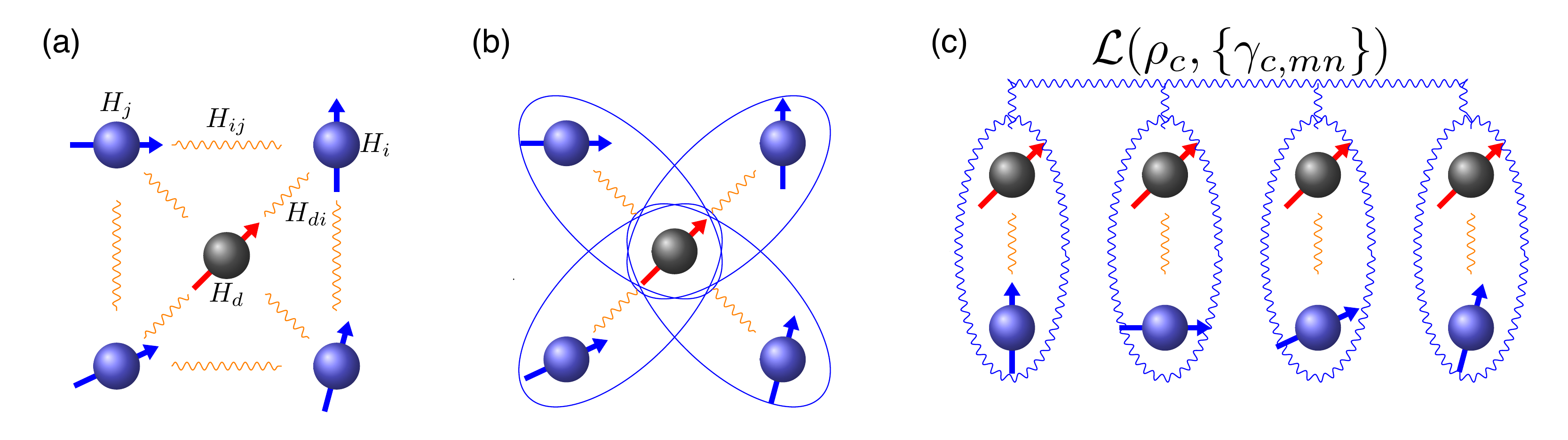}
\caption{ Illustration of the approximations used in the calculation of dipolar spin relaxation (a) Many-spin system. (b) Cluster approximation and central spin approximation of the many-spin system. (c) Coupled Lindbladian master equation to retain relevant properties of the modelled system.
\label{fig:method}  }
\end{center}
\end{figure}

In order to address this computationally highly demanding problem,  various cluster expansion\cite{yang_quantum_2008,seo_quantum_2016} and cluster approximation models have been developed.  In the calculation of the coherence function of a many-spin system, the cluster correlation expansion (CCE) method has gained considerable popularity in the recent years due to its computational efficiency and accuracy.\cite{yang_quantum_2008,seo_quantum_2016} On the other hand,  to quantitatively include spin relaxation effects one needs to go even beyond the standard CCE method.\cite{onizhuk_probing_2021} For a qualitatively accurate  description of spin relaxation a modified methodology is needed, which was developed in Ref.~[\cite{IvadyPRb2020}]. In this theory,  a cluster approximation is utilized to reduce the Hilbert space of the many-spin system.   The central spin and a few environmental spins are included in each cluster system,  see Fig.~\ref{fig:method}(b).   In contrast to the CCE method\cite{yang_quantum_2008,seo_quantum_2016},  the sub-clusters  of smaller order are not considered here.  The ensemble of these small cluster systems are meant to model the dynamics of the many-spin system.  Due to the truncation of the Hilbert space, correlation among the environmental spins are neglected.  This approximation is valid either when the coherence time of the bath spins is short or when the coupling between spins are weak, such that the intra-bath correlation cannot have an effect on the dynamics of the system in reasonable time scales.  The former is valid, for example,  when the qubit is coupled to a bath of electron spin defects with short coherence time, while the latter is valid for a weakly coupled nuclear spin bath.  Note,  that by increasing the size of the cluster systems, i.e.\ including an increasing number of environmental spins, the exact behaviour may be  asymptotically recovered. 

As can be seen, the cluster approximation naturally fits to the problem of dipolar spin relaxation, however, the dynamics of a system of independent cluster systems cannot properly describe the diagonal elements of the central spin's density matrix in the many spin system.  The most obvious reason is that the total magnetization of the cluster of independent smaller systems is not preserved.  Furthermore,  since the central spin is included in all the cluster systems,  multiples of its magnetization give rise to a false thermal state.  In order to overcome these limitations,  the dynamics of the cluster systems are bound together in Ref.~[\cite{IvadyPRb2020}] through a coupled master equation, see Fig.~\ref{fig:method}(c), which can be written in the simplest case as
\begin{equation}
 \varrho_{i}  = - \frac{i}{\hbar} \left[ \mathcal{H}_i, \varrho_i \right] + \mathcal{L} \! \left(  \varrho_i \right)  \text{,}
 \label{eq:master}
\end{equation}
where $ \varrho_{i}$ is the density matrix of cluster system $i$,  $\mathcal{H}_i = H_d + H_i + H_{di}$ is the Hamiltonian of the cluster system including the Hamiltonian of the defect spin ($H_0$) and one coupled environmental spin ($H_i$) and their interaction ($H_{di}$), and $\mathcal{L} \! \left(  \varrho_i \right)$ is an extended Lindbladian that realizes the coupling between the cluster systems.   The last term on the right hand side of Eq.~(\ref{eq:master})  can be expressed as
\begin{equation}
\mathcal{L} \! \left(  \varrho_i \right) = \sum_{j=1,j\neq i}^{N} \sum_l  \frac{\Delta a_{jl}}{\text{Tr} \! \left( C_{l}^{\dagger} C_{l} \varrho_i \right)} \left( C_l   \varrho_i C_l^\dagger - \frac{1}{2} \left( \varrho_i C_l^\dagger C_l   + C_l^\dagger C_l  \varrho_i  \right) \right) \text{,}
\end{equation}
where $C_l$ form a complete set of jump operators that can realize any population transfer between the diagonal elements of the reduced density matrix of the central spin and $\Delta a_{jl}$ are the time dependent rates that measure spin flip-flops  caused by environmental spin $j$ in the corresponding cluster system.  

The Lindbladian coupling is defined so that the diagonal elements of the reduced density matrix of the central spin are identical in all cluster systems and the global magnetization is preserved at any time during the simulation, see Ref.~[\cite{IvadyPRb2020}] for further details. This way the decaying dynamics of the coupled cluster system approaches a proper thermal state.  Relaxation of the central spin originates from the coupling to the spin bath and the decoherence of the bath.\cite{IvadyPRb2020} Decoherence properties of the bath thus have a quantitative effect on the relaxation properties.   Consequently,  the method is most accurate when the cluster approximation,  and resultant reduction of the Hilbert space,  is suitable for the considered system.  As we discussed before,  cluster approximation expectedly holds for electron spin environment of short coherence time and nuclear spin environments of weak couplings. When a spin bath includes more than one strongly coupled nuclear spins, inclusion of intra-spin-bath correlation is required, which can be achieved by increasing the number of bath spins within the cluster systems.

\subsection{Silicon carbide,  the divacancy qubits, and relevant spin defects }
\label{sec:model}

Silicon carbide is a polytypic material with the 4H polytype being the most stable\cite{kawanishi_effect_2016} and most commonly used. The primitive cell of 4H-SiC consists of four Si-C double layers in hexagonal stacking order. Due to the stacking of the double layers, there are two non-equivalent positions for single site point defects in 4H-SiC. The next nearest neighbor sites of the so-called $h$ and $k$ defect configurations show hexagonal-like and cubic-like local arrangements, respectively. In a perfect 4H-SiC lattice, the divacancy, as an adjacent pair of vacancies, has four distinguishable configurations $hh$, $kk$, $hk$, and $kh$, where the first and second letter indicate the carbon and silicon vacancy sites, respectively. These configurations in the above specified order were assigned\cite{davidsson_first_2018} to the PL1, PL2, PL3, and PL4 photoluminescence (PL) signals\cite{koehl_room_2011}. Here, we note that there are additional divacancy related PL lines observed in experiment that were named as PL5-PL6\cite{koehl_room_2011,falk_polytype_2013} as well as PL5'-PL6' and PL7-PL10\cite{MagnussonPRB2018}. Recently, some of these lines were tentatively assigned to stacking fault-divacancy complexes.\cite{ivady_stabilization_2019} In this article, we consider the four regular divacancy configurations (PL1-4) only.

There are several spin defects in 4H-SiC that can influence the spin dynamics of a point defect qubit. Considering the total spin momentum and the kind of the spins, we identify four groups of spin defects that we consider separately in this article: 1) Spin-1/2 nuclear spins due to paramagnetic $^{13}$C (1.07\%) and $^{29}$Si (4.68\%) isotopes, which are naturally abundant in SiC. 2) Dopants and intrinsic defects with a spin-1/2 electronic ground state are the dominant electron spin defects in SiC. Besides these two common spin defects, we consider two, more exotic categories, namely, 3) nearby divacancies as environmental spin defects, and 4) spin-3/2 silicon vacancies as environmental spin defects. While these defects have generally low concentrations ( $<10^{14}$~cm$^{-3}$) in high purity samples\cite{nagy_high-fidelity_2019}, due to irradiation and ion implantation, often used to create divacancy and silicon vacancy qubits, their concentration may be considerably enhanced either throughout the sample or locally. For example, after irradiation and annealing, the silicon vacancy concentration can be in the $10^{16}$ cm$^{-3}$ range or even more in heavily irradiated samples.\cite{kasper_influence_2020}

In the second group of paramagnetic defects, there are several spin-1/2 dopants and defects that can be present in SiC with varying concentrations depending on the growth process and after-growth treatments.  In as-grown high purity SiC epilayers, the dominating impurities and defects with non-zero electron spin are: a) nitrogen substitutional defects with typical concentration of $10^{14}$-$10^{15}$~cm$^{-3}$ for commercial layers, $10^{13}$~cm$^{-3}$ for ultra-pure epilayers\cite{nagy_high-fidelity_2019}, $10^{14}$-$10^{15}$~cm$^{-3}$ for high-purity semi-insulating (HPSI) 4H-SiC.\cite{Son2007} It is worth mentioning that the nitrogen nucleus also possesses a spin of $I = 1$ (99.63\% natural abundance) that couples through an isotropic hyperfine interaction of 50.97~MHz to the electron spin.\cite{greulichweber_epr_1997} b) Shallow boron defect with $10^{14}$~cm$^{-3}$ concentration for commercial layers, $10^{13}$~cm$^{-3}$ concentration for pure layers, and $10^{14}$-$10^{15}$~cm$^{-3}$ for HPSI 4H-SiC.\cite{Son2007} While boron also possesses a none zero nuclear spin, the spin density of the defect is localized on other neighboring atoms with low paramagnetic isotope abundance\cite{greulichweber_epr_1997}, thus hereinafter we neglect the hyperfine interaction for the shallow boron acceptor. c) Carbon vacancy with $10^{13}$~cm$^{-3}$ in pure epilayers and $10^{15}$~cm$^{-3}$ in HPSI 4H-SiC.\cite{nagy_high-fidelity_2019} d) Carbon antisite-vacancy (CAV) defects can be present in HPSI materials with concentration $10^{14}$-$10^{15}$~cm$^{-3}$.\cite{Son2007} e) In p-type samples the concentration of aluminum is typically in the range of 1-5$\times 10^{18}$~cm$^{-3}$. The hyperfine splitting due to paramagnetic aluminum isotope is not resolved in 4H-SiC.\cite{greulichweber_epr_1997}

\section*{Results}
\label{sec:res}

\subsection{Resonant spin bath couplings}
\label{sec:cou}

\begin{figure}[h!]
\begin{center}
\includegraphics[width=0.55\columnwidth]{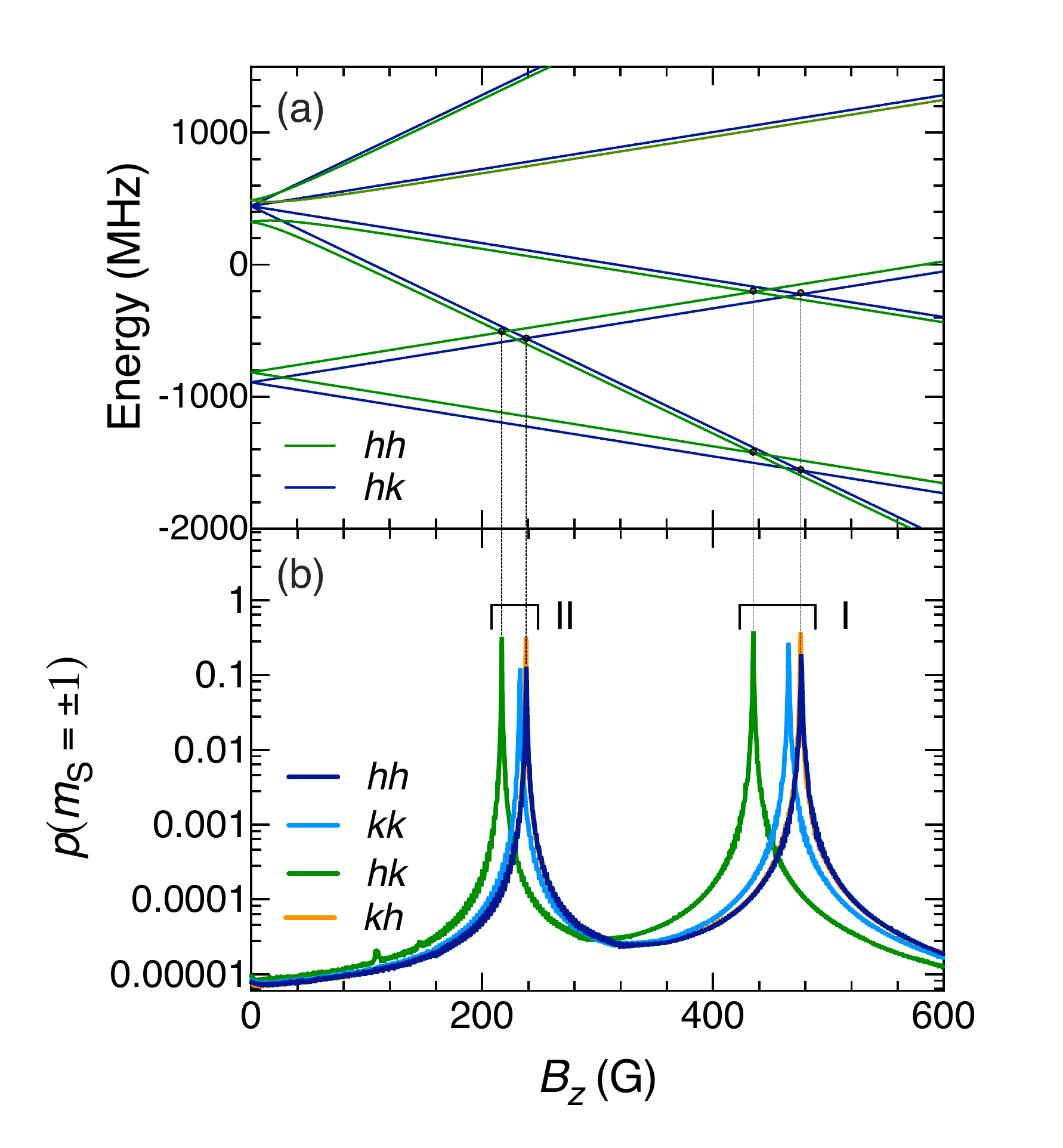}
\caption{  a) Magnetic field dependence of the  energy levels of $hh$ (dark blue) and $hk$ (green) divacancies coupled to a single spin-1/2 point defect. b) Population of the $m_S = \pm 1$ spin states of the $hh$ (dark blue), $kk$ (light blue), $hk$ (green), and $kh$ (amber) divacancies induced by a bath of spin-1/2 environmental defects at the concentration of 10$^{18}$ cm$^{-3}$. The depicted curves show the magnetic field dependence  of the population enhancement of the  $m_S = \pm1$ states of the divacancies. Each curve exhibits two resonances corresponding to $B_{\text{GSLAC}} \! \left( D, E \right)$ (group I) and $B_{\text{GSLAC}} \! \left( D, E \right)/2$ (group II), where $D$ and $E$ are the ZFS parameters of the corresponding divacancy qubit. In a) and b) the magnetic field is parallel to the quantization axis of the divacancy qubits in all cases. At $t = 0 $, the central divacancy is initialized in the $m_S = 0$ state, while the spin bath in an high temperature thermal state.
\label{fig:spin-12e}  }
\end{center}
\end{figure}

First we study resonance effects due to the most common spin defects in SiC, i.e. nuclear spin and spin-1/2 point defects. Spin mixing curves due to $^{29}$Si and $^{13}$C nuclear spins exhibit a single resonance place that corresponds to ground state level anticrossing (GSLAC) of the divacancy electron spin states.  At the magnetic field value of the GSLAC, $B_{\text{GSLAC}} \! \left( D , E \right) = \left( D-E \right) / g_e \mu_B $ the Zeeman shift of the $m_S = -1$ divacancy spin states compensates the zero-field-splitting. The gapless electron spin states can be efficiently mixed either with each other through the precession of the electron spins due to transverse magnetic field, or with nuclear spin states due to the hyperfine interaction. The latter effect gives rise to strong relaxation effect at $B_{\text{GSLAC}}$. Note that for different divacancy configurations the GSLAC resonance appears at different magnetic field values due to the different $D$ and $E$ values of the centers. Note also that the different divacancy configurations possess different neighboring shells, which, however, on average has minor effects  on the overall strength of the mixing.

\begin{figure}[h!]
\begin{center}
\includegraphics[width=0.65\columnwidth]{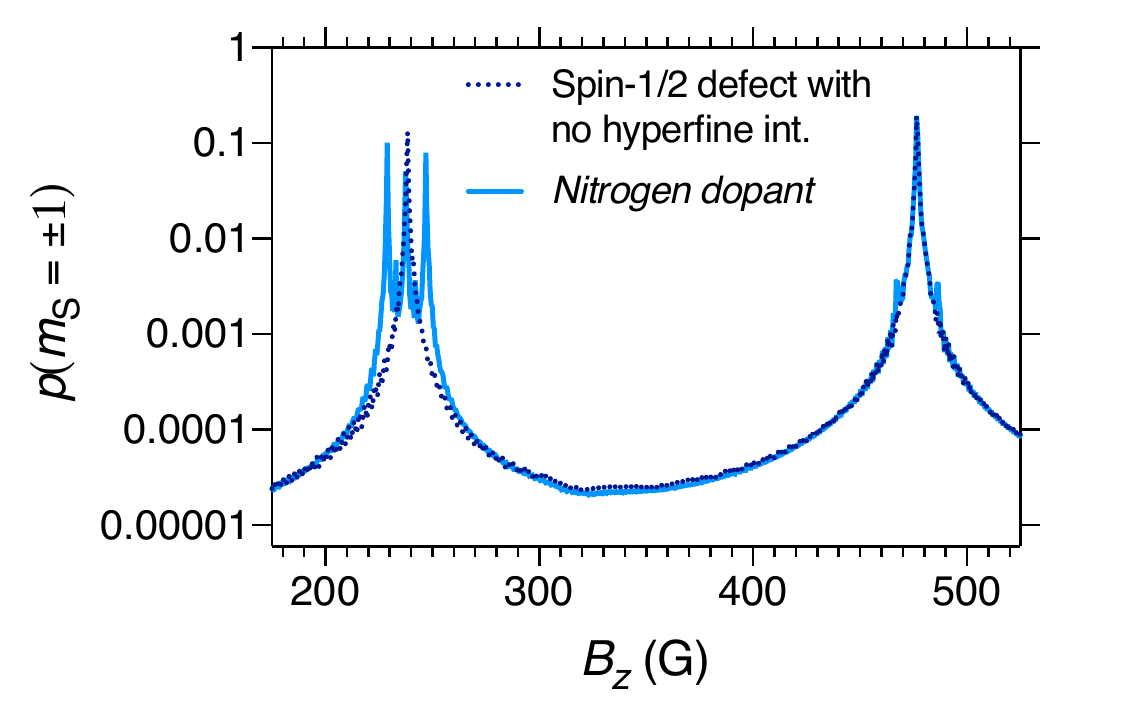}
\caption{  Population of the $m_S = \pm 1$ spin states for the case of spin-1/2 defect with strong hyperfine interaction. Solid light blue line shows the population variation of the $hh$ divacancy spin states due to substitutional nitrogen shallow donor. Dotted dark blue depicts population variation due to spin-1/2 electron spin defects without the hyperfine coupling. Hyperfine interaction leads to  a splitting of the center peak into three peaks at $B_{\text{GSLAC}}/2$ as well as to an appearance of two satellite peaks next to the GSLAC peaks.  
\label{fig:spin-12e+i}  }
\end{center}
\end{figure}

Spin-1/2 point defects, like boron, aluminum, nitrogen dopants, carbon vacancies, and CAV defects give rise to two distinct level crossings at $B_{\text{GSLAC}}$ and $B_{\text{GSLAC}}/2$, see Fig.~\ref{fig:spin-12e}(a), that result in resonant spin relaxation, see group I and II peaks in Fig.~\ref{fig:spin-12e}(b). The GSLAC resonance is due to the precession of the divacancy spins induced by the transverse field of the spin defects. At $B_{\text{GSLAC}}/2$, the $\left| m_S, m_{\mathcal{S},i} \right\rangle = \left| 0, +1/2 \right\rangle$  and $ \left|-1, -1/2 \right\rangle$  states cross. Due to the dipole-dipole interaction of the neighboring paramagnetic defects, a LAC is formed at $B_{\text{GSLAC}}/2$ and efficient spin flip-flop interaction takes place. While divacancies can polarize nuclear spin at $B_{\text{GSLAC}}$, they can polarize spin-1/2 electron spins only at $B_{\text{GSLAC}}/2$. Note that when a strong hyperfine coupling is not expected at the spin defect site, a featureless resonance peak is observed as seen in Fig.~\ref{fig:spin-12e}(b). On the other hand, the nitrogen dopant, exhibits a strong hyperfine interaction with its abundant $^{15}$N nuclear spin, resulting in a fine structure, see the dashed line in Fig. 2 for the $hh$ divacancy. The hyperfine interaction at the spin defect site leads to a splitting of the resonance peaks in group II at $B_{\text{GSLAC}}/2$ according to the nuclear spin states. In addition, two satellite peaks appear on the GSLAC resonance peak, see Fig.~\ref{fig:spin-12e+i}. The curves are similar to those reported for the NV center in diamond interacting with substitutional nitrogen point defects \cite{WickenbrockAPL2016,ivady_photoluminescence_2021}. However, in hexagonal SiC polytypes there is only one allowed symmetry distortion and the hyperfine interaction is dominantly isotropic\cite{greulichweber_epr_1997}, which leads to a less complicated spin structures at $B_{\text{GSLAC}}/2$.

\begin{figure}[h!]
\begin{center}
\includegraphics[width=0.5\columnwidth]{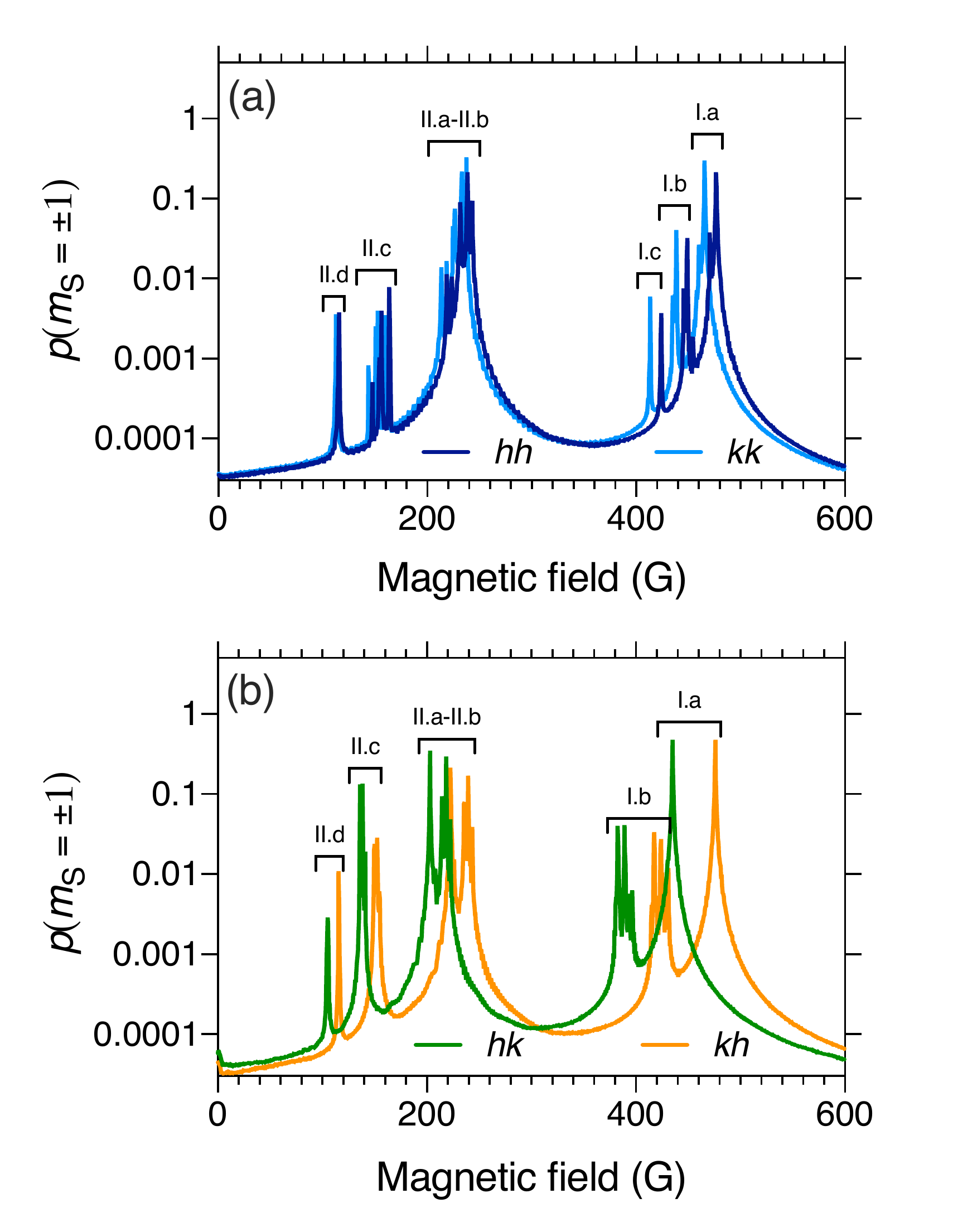}
\caption{  Magnetic field dependence of the population of the divacancy $m_S = \pm 1$ spin states due to spin flip-flops induced by V1 and V2 silicon vacancy centers. a) and b) depict the magnetic field dependence of the spin mixing of $hh$ and $kk$ and $hk$ and $kh$ divacancy centers, respectively.  In addition to the  primary peaks I.a and II.a, observed for spin-1/2 defect, satellite peaks I.b, I.c, II.b, and II.c can be observed to the left of their respective primary peaks. The satellite peaks correspond to double and triple quantum jumps of environmental silicon vacancies.  In a) and b) the magnetic field is parallel to the quantization axis of the divacancy qubits in all cases.  At $t = 0 $, the central divacancy is initialized in the $m_S = 0$ state, while the spin bath in an high temperature thermal state.
\label{fig:spin-32e}  }
\end{center}
\end{figure}

\begin{figure}[h!]
\begin{center}
\includegraphics[width=0.5\columnwidth]{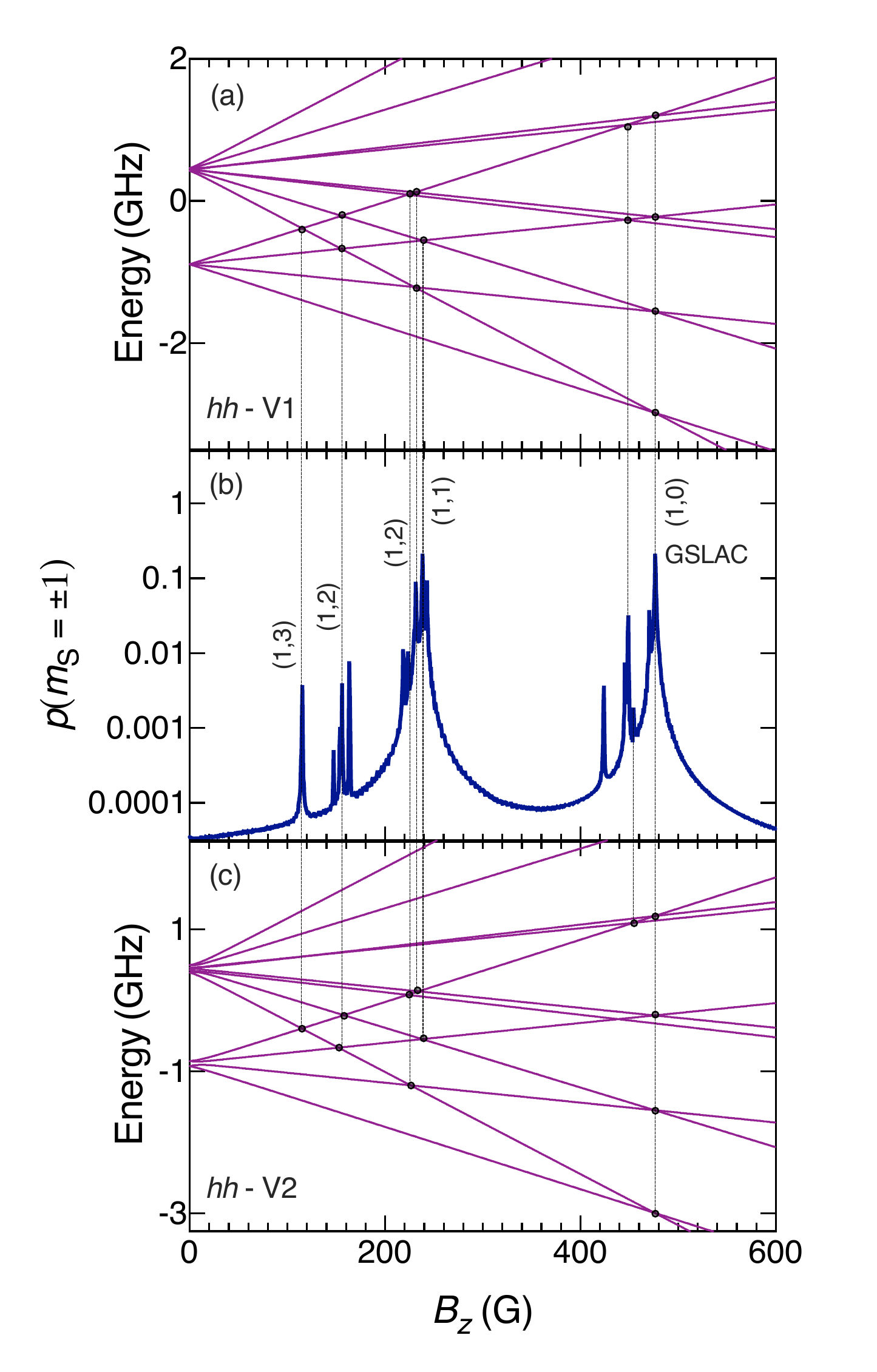}
\caption{  Identification of resonance peaks for the $hh$ divacancy interacting with spin-3/2 silicon vacancy centers. a) and c) show the energy levels structure of $hh$ divacancy-V1 silicon vacancy and $hh$ divacancy-V2 silicon vacancy two spin systems, respectively. b) depicts the derived population curve for $hh$ divacancy. At each resonance peak, a pair of numbers in parenthesis provide the jump of the spin quantum numbers of the divacancy and silicon vacancy centers at the assigned level crossings. 
\label{fig:spin-32e-det}  }
\end{center}
\end{figure}

More exotic couplings are possible when the divacancies interact with other divacancies or with spin-3/2 silicon vacancy centers. This may be the case in highly irradiated samples or when the sample is bombarded with positive ions creating hundreds of defects in a small volume\cite{DefoPRB2018}. In Fig.~\ref{fig:spin-32e}, we depict the magnetic field dependence of the population of the divacancy $ m_S = \pm 1$ spin states due to spin flip-flops induced by V1 and V2 silicon vacancy centers\cite{IvadyVSi-4H}. The curves resemble in their main character the mixing curves of the spin-1/2 defects. It is indeed expected as the quartet spin of the silicon vacancy contains a $m_{\mathcal{S}} = \left\lbrace +1/2, -1/2 \right\rbrace$ manifold that can interact with the divacancy spin states as the spin-1/2 defect spins. Accordingly, we observe two main resonance I.a and II.a at magnetic field values $B_{\text{GSLAC}}$ and $B_{\text{GSLAC}}/2$. In addition to these dominant peaks several satellite peaks can be observed on the left side of both primary resonances. Majority of the peaks can be connected to level crossings, see Fig.~\ref{fig:spin-32e-det}. To allow spin mixing at these crossings the quantum number of the silicon vacancy spin must jump either by two or by three, see Fig.~\ref{fig:spin-32e-det}. Note that dipole-dipole interaction does not include such Hamiltonian terms. However, higher order Zeeman terms are non-zero for quartet spin states in C$_{3v}$ symmetry, see in Eq.~(\ref{eq:VSIZ}),  which in combination with the dipole-dipole interaction makes a weak mixing of the divacancy and silicon vacancy spin states possible. Peaks II.d require the triple quantum jumps, i.e. ($\Delta m_S, \Delta m_\mathcal{S}$) = (1,3), while peaks I.b, II.b, and II.c require double quantum jumps from the silicon vacancy spin, i.e. ($\Delta m_S, \Delta m_\mathcal{S}$) = (1,2).

\begin{figure}[h!]
\begin{center}
\includegraphics[width=0.6\columnwidth]{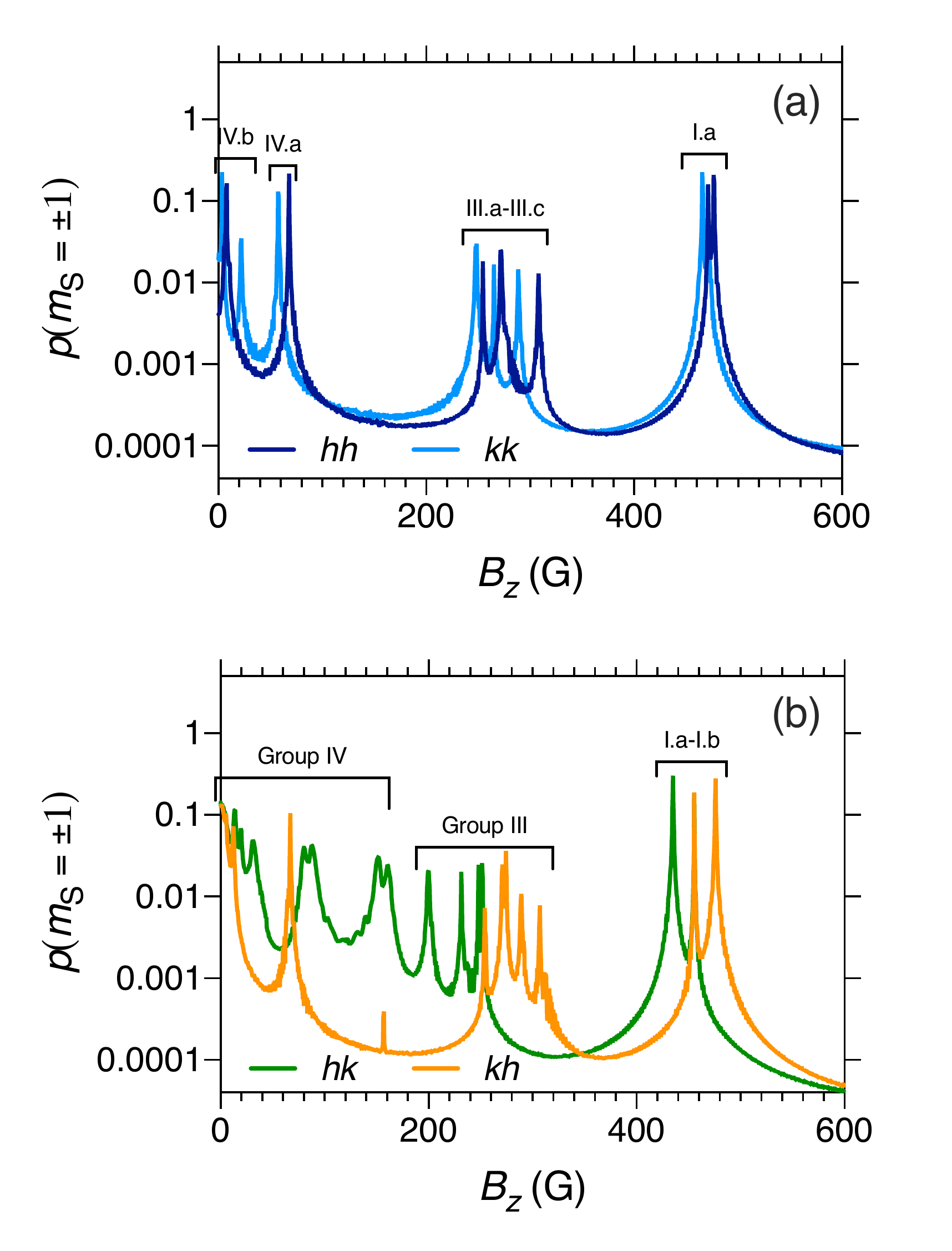}
\caption{  Population of the $m_S = \pm 1$ spin states of divacancies interact with other divacancy spins. a) and b) show the mixing curves on a logarithmic scale of the $hh$ and $kk$ and the $hk$ and $kh$ divacancy centers, respectively. In all cases the spin bath contains all the possible divacancy configurations other than the central divacancy under consideration. In a) and b) the magnetic field is parallel to the quantization axis of the divacancy qubits in all cases.  At $t = 0 $, the central divacancy is initialized in the $m_S = 0$ state, while the spin bath in an high temperature thermal state.
\label{fig:spin-1e}  }
\end{center}
\end{figure}

\begin{figure}[h!]
\begin{center}
\includegraphics[width=0.60\columnwidth]{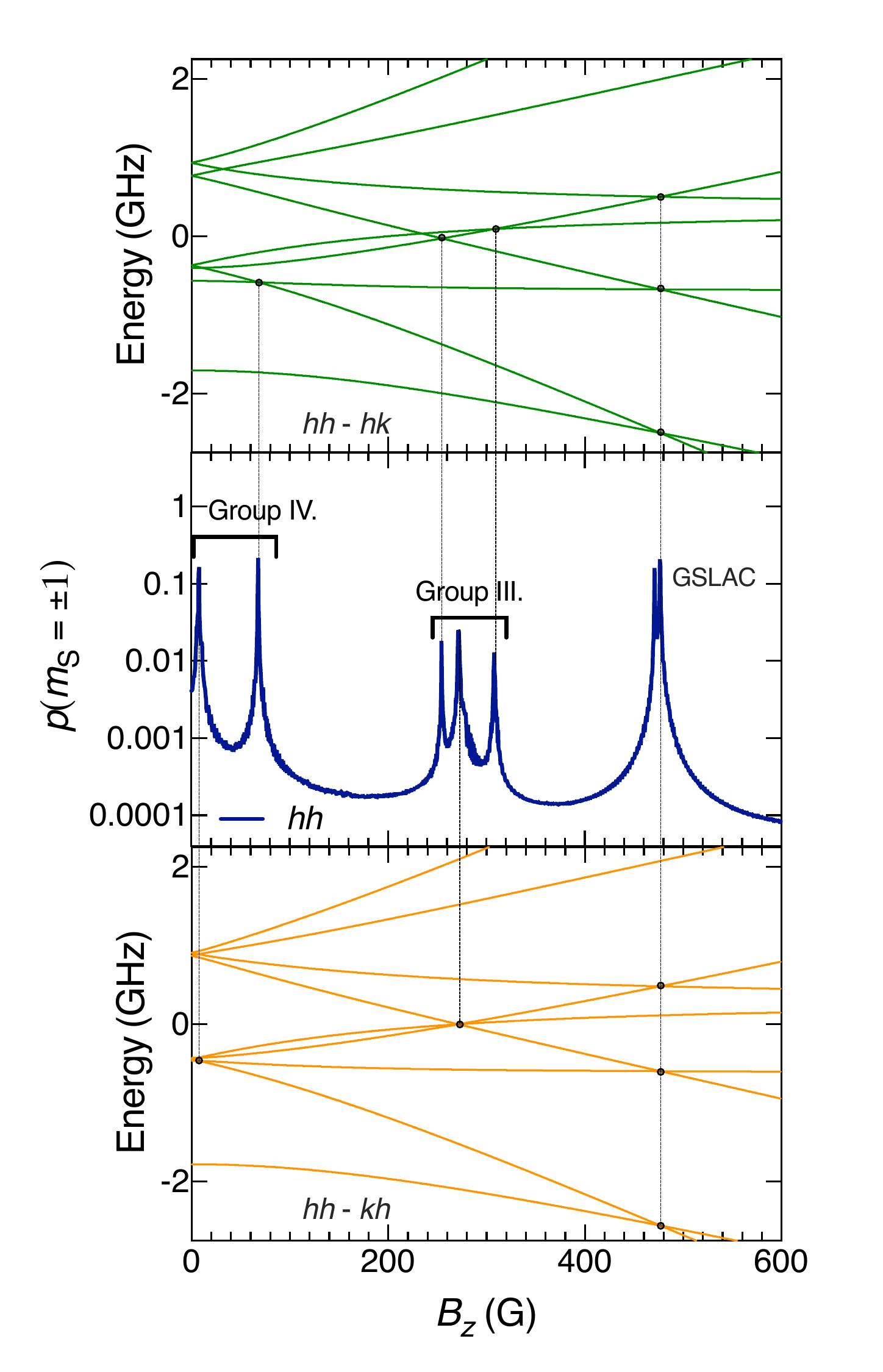}
\caption{  Identification of resonance peaks for the $hh$ divacancy interacting with other divacancy centers. a) and c) show the energy levels structure of $hh$ divacancy-$hk$ divacancy and $hh$ divacancy-$kh$ divacancy two spin systems, respectively. At zero magnetic field the spin states split into three well separated manifolds. b) depicts the derived spin population curve for the $hh$ divacancy.
\label{fig:spin-1e-id1}  }
\end{center}
\end{figure}

In order to investigate the interaction between divacancy centers we studied each divacancy configuration interacting with a bath of other possible divacancies. The obtained spin polarization curves exhibit a complicated pattern which has not been seen so far, see Fig.~\ref{fig:spin-1e}. Besides the GSLAC resonance, C$_{3v}$ symmetric configurations exhibit two other groups of resonance peaks, peaks III and IV. In order to understand the origin of these resonances, we shortly discuss the energy level structure of divacancy pairs. At zero magnetic field, the states of all divacancy pairs, irrespective to their local quantization axis, split into three manifolds, 1) the single degenerate $\left| m_S, m_{\mathcal{S}} \right\rangle = \left| 0, 0 \right\rangle$ manifold with the lowest energy, 2) the  $\left\lbrace  \left| 0, \pm 1 \right\rangle, \left| \pm 1, 0 \right\rangle \right\rbrace$ manifold, and C) the $ \left\lbrace \left|\pm1, \pm1 \right\rangle, \left| \mp 1 , \pm 1 \right\rangle  \right\rbrace $ manifold with the highest energy, see Fig.~\ref{fig:spin-1e-id1}. The GSLAC resonance is due to the crossing between the $ \left| 0, 0 \right\rangle$ state and states of the $\left\lbrace \left| 0, \pm1 \right\rangle, \left| \pm 1, 0 \right\rangle \right\rbrace$ manifold. Group III peaks appear at crossings of manifolds 2) and 3), see Fig.~\ref{fig:spin-1e-id1}. Note that spin mixing at such level crossings is only possible when divacancies with two different quantization axes are interacting. In this case the Zeeman interaction partially mixes the spin states of one of the divacancies. Combined with dipole-dipole interaction this makes spin flip-flop interaction possible at the crossings. Accordingly, peaks III.a and III.c and peak III.b in Fig.~\ref{fig:spin-1e}(a) originate from interactions with $hk$ and $kh$ basal plane oriented divacancy centers, respectively, see Fig.~\ref{fig:spin-1e-id1}. Group III peaks include four peaks for basal plane oriented divacancy configurations (Fig.~\ref{fig:spin-1e}(b)), each of them corresponds to interaction with divacancies, $hh$, $kk$, $hk$, and $kh$, whose quantization axis is 109~degree aligned to the quantization axis of the considered basal plane divacancy. Group IV. peaks correspond to crossings between the states of the $\left\lbrace \left| 0, \pm1 \right\rangle, \left| \pm 1, 0 \right\rangle \right\rbrace$ manifold, see Fig.~\ref{fig:spin-1e-id1}. As the splitting of the states in the middle manifold is small compared with the $D$ zero-field-splitting parameter, the resonances peaks appear at small magnetic field values in most cases.  The positions of the crossings largely depended on the differences of the $D$ splittings and the values of the $E$ splitting of the coupled divacancy centers. The ZFS parameters of the $hk$ divacancy deviates the most from the ZFS parameters of the other configurations. Consequently, the crossing region stretches over a magnetic field interval of 175~G. Furthermore, accidental degeneracies can be observed in the energy level structure (not shown) that give rise to broad resonance peaks as can be seen in Fig.~\ref{fig:spin-1e}(b) for the $hk$ divacancy center.

As the different spin defects give rise to different signatures, study of spin mixing can be utilized to identify the dominant spin-spin interactions in a sample. In case of single defects, spin mixing, detectable through the spin dependent luminescence of the centers, can provide valuable information on the local environment of the centers.

\subsection{Longitudinal spin relaxation}
\label{sec:lon}

\begin{figure}[h!]
\begin{center}
\includegraphics[width=0.6\columnwidth]{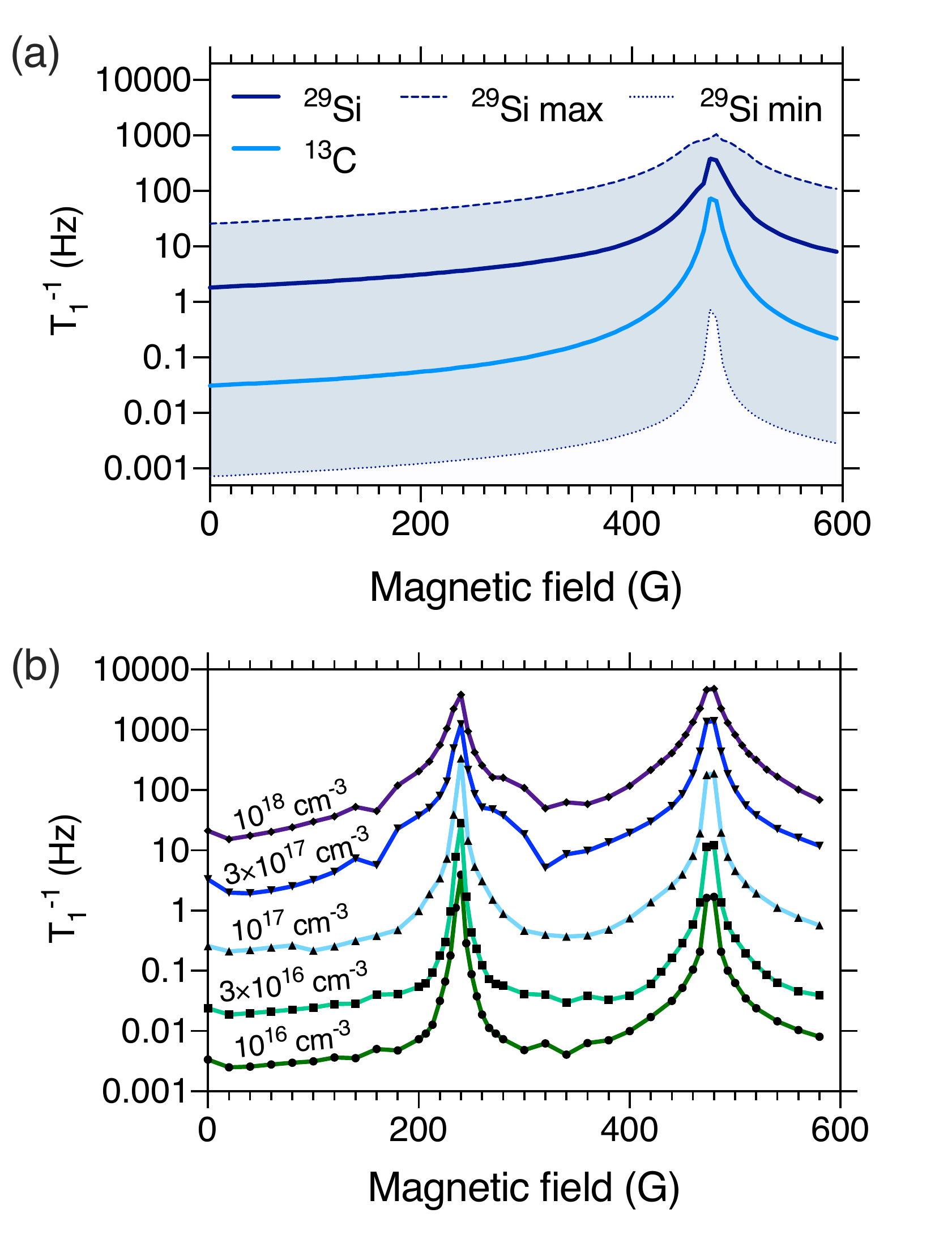}
\caption{  Magnetic field dependence of the longitudinal spin relaxation rate T$_1^{-1}$ due to a) nuclear spin of natural abundance and b) spin-1/2 point defects of various concentrations. a) Relaxation due to $^{29}$Si (solid dark blue line) and $^{13}$C (solid light blue line) nuclear spins are considered separately. When $^{13}$C spin bath is considered the nearest neighbor sites are excluded. For $^{29}$Si we also depict the largest (dashed dark blue line) and the smallest (dotted dark blue line) relaxation rate found in our ensemble of 100 configurations. The shaded are between these curves covers more than four orders of magnitude.
\label{fig:longi}  }
\end{center}
\end{figure}

In this section, we carry out quantitative study on the relaxation dynamics of the divacancy qubits in different spin environments and determine the magnetic field and concentration dependence the longitudinal spin relaxation rate T$_1^{-1}$. Here, we focus only on the most relevant spin defects in SiC, namely $^{29}$Si and $^{13}$C nuclear spins and spin-1/2 point defects. Figs.~\ref{fig:longi} a) and b) depict the calculated spin relaxation rate for $hh$ divacancy qubit embedded in a nuclear spin environment and spin-1/2 point defect environment of different concentrations, respectively. Here, we note that when $^{13}$C spin bath is considered the nearest neighbor sites are excluded. Comparing the $^{13}$C nuclear spin environment induced spin relaxation rate with the one obtained for the NV center\cite{IvadyPRb2020}, one can notice that the paramagnetic carbon nuclei of natural abundance give rise to comparable but somewhat smaller relaxation rate for the $hh$ divacancy in SiC. This is due to the fact that the lattice constant of SiC is larger than in diamond, thus the hyperfine interaction decreases faster within the neighboring shells. In addition, carbon atoms occupy only half of the sites in the vicinity of the defect. For example, the second neighborhood shell, that contains twelve sites with hyperfine interaction $\approx$10~MHz, is occupied by silicon atoms. This very last argument suggests that $^{29}$Si paramagnetic silicon nuclei of 4.6\% natural abundance have strong effect on the spin relaxation time.  Indeed, according to our simulations $^{29}$Si nuclear spins limit the ensemble averaged T$_1$ time to 100-500~ms farther away from the GSLAC resonance. We note that the spin relaxation rate of individual divacancy centers can vary over several orders of magnitude due to their distinct local environment. As can be seen in Fig.~\ref{fig:longi}, the largest relaxation rate found in our ensemble is an order of magnitude larger than the average, while the smallest one is more than three orders of magnitude smaller than the average.

Comparing Figs.~\ref{fig:longi}(a) and (b), one can see that, spin-1/2 point defect induced relaxation starts to dominate, except the resonance place at $B_{\text{GSLAC}}/2$, only at defect concentrations as high as $3 \times 10^{18}$ cm$^{-3}$. Therefore, we conclude that in samples of natural isotope abundance, $^{29}$Si nuclei are the main non-thermal source of spin relaxation. In $^{29}$Si isotope depleted samples, depending on their net concentration, spin-1/2 point defects may dominantly contribute to spin relaxation at low temperature. Fig.~\ref{fig:longi}(b) shows the magnetic field dependence of the relaxation rate at various concretions of spin-1/2 point defects ranging from 10$^{16}$ cm$^{-3}$ to 10$^{18}$ cm$^{-3}$. Note that the lowest concentration considered in this study is still considered to be high in SiC, however, below this level $^{13}$C nuclear spins start to dominate the relaxation dynamics.

\begin{figure}[h!]
\begin{center}
\includegraphics[width=0.65\columnwidth]{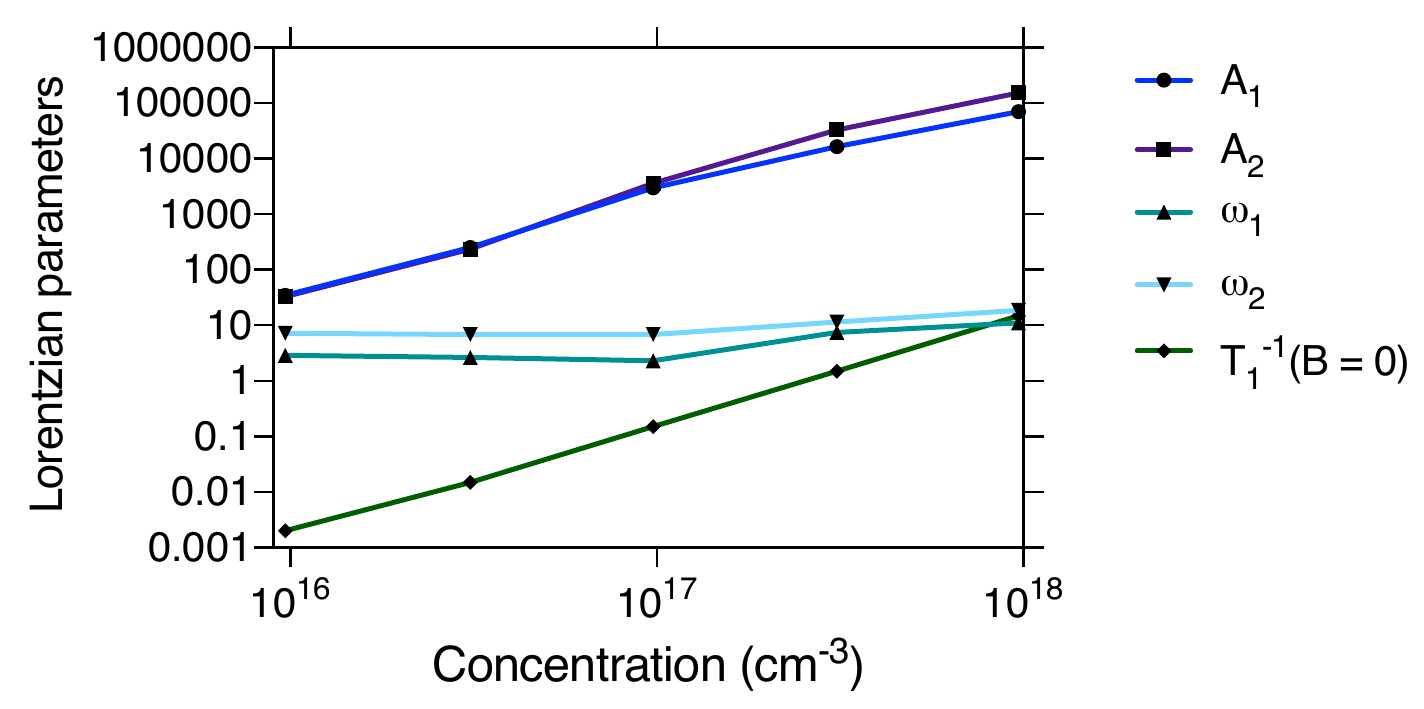}
\caption{  Concentration dependence of the fitting parameters of Eq.~(\ref{eq:dlor}) fitted to the simulated spin-1/2 defect concentration dependent spin relaxation rate curves in Fig~\ref{fig:longi}(b).  
\label{fig:conci}  }
\end{center}
\end{figure}

In order to make our numerical results more accessible, we parameterize all the curves depicted in Fig.~\ref{fig:longi}(b).  A Lorentzian curve is adequate to describe the effects of resonant spin state couplings in the few spin limit.  Due to the large number of spins included in the bath and the vast variation of spin bath configurations in an ensemble,  the actual spin relaxation curve at the resonance places can be considered as a superposition of numerous Lorentzian curves of varying width.  The distribution of the width of the Lorentzian curves  is dictated by the distribution of the coupling strength between the defect and the spins in the environment.  For simplicity,  we fit a single Lorentzian at each resonance peak that can describe the decay of the relaxation rate properly farther away from the resonance place, but may be inaccurate right at the place of the level crossing.  The relaxation rate is thus parameterized as
\begin{equation} \label{eq:dlor}
T_1^{-1} \! \left( B, C \right) = T_1^{-1}\! \left( 0, C \right)  + \frac{2A_1}{\pi}  \frac{\omega_1}{ 4 \left( B - B_{\text{GSLAC}/2} \right)^2 + \omega_1^2 } + \frac{2A_2}{\pi}   \frac{\omega_2}{ 4 \left( B - B_{\text{GSLAC}} \right)^2+\omega_2^2 } \text{,}
\end{equation}
where $B$ is the magnetic field, $C$ is the concentration of spin-1/2 defect, $B_{\text{GSLAC}/2} \approx 240$~G and  $B_{\text{GSLAC}} \approx 480$~G.  $T_1^{-1}\! \left( 0, C \right)$, the relaxation rate at $B=0$, $A_1$, $A_2$, $\omega_1$, and $\omega_2$ are fitting parameters which are depicted in Fig.~\ref{fig:conci} as a function of the spin-1/2 defects concentration. In order to be able to extrapolate to concentrations that are not considered in our numerical simulations, we fitted polynomial curves to the concentration dependence of the parameters of Eq.~(\ref{eq:dlor}), as
\begin{equation} \label{eq:parc}
X\! \left( C \right) = \alpha + \beta C^n \text{,}
\end{equation}
where $ X$ is a parameter of the double-Lorentzian curve. The simple polynomial form given in Eq.~(\ref{eq:parc}) fits well onto the points depicted in Fig.~\ref{fig:conci}. Table~\ref{tab:par} provides the parameters of these polynomial fits. To obtain the relaxation rate for a particular defect concentration given in cm$^{-3}$ unit, one can determine the corresponding Lorentzian parameters using Eq.~(\ref{eq:parc}) and Table~\ref{tab:par} and insert them into Eq.~(\ref{eq:dlor}). This will provide the ensemble averaged relaxation rate in Hz unit.

\begin{table}[!h]
\caption{Parameters of the polynomial fits describing concentration dependence parameters of the double-Lorentzian curve fitted to the numerical results.}
 \begin{center}
       \begin{tabular}{c|ccc} 
             Parameters of the Lorentzians    &   $\alpha$   &  $\beta$ &  n   \\  \hline 
A$_1$ & 0 & $1.75 \times 10^{-25}$ & 1.65  \\
A$_2$ & 0 & $1.23 \times 10^{-29}$ & 1.9 \\
$\omega_1$ & 2.87 & $1.6 \times 10^{-15}$ & 0.875 \\
$\omega_2$ & 6.89 & $1.23 \times 10^{-17}$ & 1 \\
T$_1^{-1} \! \left( B = 0 \right)$ & 0 & $ 1.6 \times 10^{-35}$ & 2
       \end{tabular}
 \end{center}
\label{tab:par}
\end{table}

Considering the concentration dependence of the spin-1/2 point defects induced relaxation rate at zero magnetic field, we obtain a simple quadratic concentration dependence of the relaxation rate,
\begin{equation} \label{eq:t1-1}
T_1^{-1} \left( B=0, C \right) = \beta C^2
\end{equation}
where $\beta = 1.6 \times 10^{-35}$ Hz/cm$^{-6}$. Note that the quadratic dependence given by Eq.~(\ref{eq:t1-1}) differs from the earlier experimental reports\cite{jarmola_temperature_2012} observing linear dependence on the concentration. We attribute the difference to the uncertainties in the concentration and relaxation rate measurements and to the smaller concentration interval considered in the experiment.

\section*{Discussion}
\label{sec:das}

As we have seen in Eq.~(\ref{eq:rate}),  there are two main contributions to the temperature independent part of the spin relaxation,  namely the dipolar spin relaxation, due to the surrounding spin bath, and the temperature independent contribution of the direct one phonon scattering process.  Separating the contributions of these processes is sometimes cumbersome and requires low temperature measurements\cite{astner_solid-state_2018}.  As can be seen from the results of Ref.~[\cite{lin_temperature_2021}] and from our fit provided in Table~\ref{tab:parA},   direct relaxation plays an important role for divacancy defects,  thus its temperature independent contribution may not be ruled out.  On the other hand,  this process is sample independent, in contrast to the dipolar spin relaxation, as it depends only on the intrinsic properties of the SiC. Therefore,  the lowest ever reported  relaxation rate of divacancy spins in SiC sets an upper bound for the related constant $A_{\text{dir,0}}$.  So far the longest relaxation time of divacancy defects is reported in 6H-SiC and found to be in the 10~ms range at 20~K.\cite{falk_polytype_2013}  Note that 4H and 6H-SiC polytypes share common properties and we expect that the $A_{\text{dir,0}} $ parameter is in the same order of magnitude in 4H-SiC.  This assumption is further supported by the results of Ref.~[\cite{miao_universal_2020}], where the reported 64~ms coherence time indicates a relaxation time in the order of 10~ms in 4H-SiC.  This means that the $A_{\text{dir,0}} < 0.1$~kHz, which is  an order of magnitude smaller than the fitted values in Table~\ref{tab:parA}.   Consequently,  the temperature independent relaxation rate of  divacancy qubits in the nitrogen ion irradiated sample in Ref.~[\cite{lin_temperature_2021}] is dominantly due to dipolar spin relaxation processes and it is found in the range of 1.8~ms$^{-1}$ and  1.6~ms$^{-1}$ at 180~G external magnetic field for 10$^{14}$ and 10$^{13}$ cm$^{-2}$ implantation doses, respectively. 

Our results depicted in Fig.~\ref{fig:longi} clearly show that the hyperfine interaction cannot account for the 1.8~ms$^{-1}$ and  1.6~ms$^{-1}$ dipolar relaxation rate at 180~G, it is three orders of magnitude smaller.  On the other hand,  paramagnetic defects created by nitrogen ion irradiation can account for the observed values.  Utilizing the analytical fit provided in Eqs. ~(\ref{eq:dlor}) and (\ref{eq:parc}) and Table~\ref{tab:par}, we obtain $C=4.1 \times 10^{18}$~ cm$^{-3}$ and $3.9 \times 10^{18}$~ cm$^{-3}$ defect concentration for the 10$^{14}$~cm$^{-2}$ and 10$^{13}$~cm$^{-2}$ N$_2$ ion irradiated samples.  Such a high defect contraction may be expected due to the large irradiation dose applied in the experiment.  We note, however, that the concentration only slightly depends on the implantation dose.   A possible  explanation of the implantation dose independence could be given based on the fact that the T$_1$ time based concentration measurement provides information only on the local environment of the defects and not on the concentration of the entire sample.  Since positive ions create a large damage locally in an implantation layer, it is expected that the defect concentration is highly inhomogeneous.  Furthermore,  different ions create different local damage, thus the observed $C \approx 4  \times 10^{18}$~ cm$^{-3}$ defect concentrations may characterize the method of implantation (30~keV N$_2$ ions) rather than the dose and the sample's overall defect concentration.
Finally,  we note also that the selection of the external magnetic field value is quite unfortunate, since at this magnetic field the divacancy defects can couple relatively strongly to both spin-1/2 defects as well as the spin-3/2 silicon vacancy, see Fig.~\ref{fig:spin-12e} and  Fig.~\ref{fig:spin-32e-det}.

\section*{Summary}

In this article we have reported on a systematic study of the spin relaxation dynamics of divacancy qubits in 4H-SiC. First, we examined the magnetic field dependence of the spin mixing induced by various spins in the local environment of divacancy qubits. We demonstrated that  neighboring divacancy centers and the spin-3/2 silicon vacancy centers give rise to rich relaxation patterns with multiple of relaxation peaks. The different magnetic field dependent spin relaxation patterns reported in this article may allow one to identify and study the local environments of a single or an ensemble of  divacancy qubits by optical means. Related applications, such as dynamic nuclear polarization, microwave free sensing, and spin relaxation spectroscopy are yet to be explored in SiC.

Moreover, we have simulated the magnetic field and the concentration dependences of the spin relaxation time T$_1$ for the most relevant spin defects in SiC. We have shown that in high purity samples the dominant non-thermal contribution to the spin relaxation comes from the $^{29}$Si nuclear spin bath, that maximizes the ensemble averaged T$_1$ time at 100 ms at low temperature far away from the GSLAC resonance. For configurations of adjacent nuclear spins the relaxation time may reduce to 40~ms at zero magnetic field. Note that the $^{29}$Si spin bath limited spin relaxation time is  comparable or even shorter than the coherence time of 64~ms reported in Ref.~[\cite{miao_universal_2020}] for a decoherence protected subspace of divacancy qubits. In such cases, the longitudinal spin relaxation may be the major  limiting factor for the lifetime of the coherence protected subspace.

We also demonstrated that dipolar spin relaxation of divacancies due to paramagnetic point defects may be significant in ion implanted samples.  The  analytical formula provided in this paper can be utilized either to estimate T$_1$ in a given sample of known spin defect concentration or to analyze the measured T$_1$ time to estimate the local spin defect concentration of divacnacy qubits. By using this latter approach, we showed that the local concentration of paramagnetic point defect in N$_2$ ion implanted samples can be as high as $4 \times 10^{18}$~cm$^{-3}$ that maximizes the coherence time in about 0.5~ms.  This value is a factor of two smaller than the 1.3~ms coherence time achievable in 4H-SiC of natural isotope abundance.\cite{seo_quantum_2016}

\section*{Methodology}
\label{sec:met}

In order to faithfully simulate spin relaxation processes, we utilize the method developed in Ref.~[\cite{IvadyPRb2020}]. In the simulations we consider a central divacancy defect interacting with its environment, i.e. a bath of spins consisting of one type of spin defects from categories 1)-4), in a central spin arrangement. Considering only one type of spins, typically the closest spin defects play the major role in thermalizing the central spin. We found that inclusion of the 32 closest defects already provides convergent results in terms of spin relaxation.  The many-spin system is then divided into a cluster of subsystems which includes the central divacancy and one defect spin each, see Fig.~1 in Ref.~[\cite{IvadyPRb2020}]. The applied theoretical method\cite{IvadyPRb2020} introduces an effective interaction between the subsystems in such a way that the total spin of the many-spin system is preserved throughout the simulation. This ensures reliable simulation of spin relaxation and calculation of T$_1$, as demonstrated for the case of the NV center in Ref.~[\cite{IvadyPRb2020}]. In this article, we study ensemble averaged quantities. Therefore, in all cases an ensemble of randomly generated local spin environments is considered. In our qualitative (quantitative) study of divacancy ensembles we consider 100 (200) random spin configurations corresponding to a given defect concentration on average. The time step is set to 25~ps (20~ps), while the simulation time is set to 1~$\mu$s (600~$\mu$s) for the qualitative (quantitative) analysis of the relaxation processes. When the hyperfine interaction induced spin relaxation is investigated, 100~ps time step and 600~$\mu$s simulation time are used, and ensemble averaging is carried out over 100 random configurations of a bath containing 128 nuclear spins.  Increasing the number of spins is motivated by the spatial extension of the spin density of the divacancies, which is comparable with the distances of the closest spins alike to the case of point defect spins.

In our real time simulations we investigated the time dependent variation of the initial population of the divacancy spin states. At $t = 0 $, the central divacancy is initialized in the $m_S = 0$ state, while the spin bath in an high temperature thermal state. 

We simulate the dynamics of the coupled spin system without any additional approximation on the local Hamiltonian of the divacancy, the spin defects, and the environmental coupling of the divacancy. Intra-spin-bath interactions, on the other hand, are neglected. The spin Hamiltonian thus can be written in the form of
\begin{equation}
H = H_{\text{div}} + H_{\text{bath}} + H_{\text{coup}} \text{,}
\end{equation}
where the Hamiltonian $ H_{\text{div}} $ of the divacancy consists of three terms
\begin{equation} \label{eq:div}
H_{\text{div}} = g_e \mu_{\text{B}} B_z S_z + D \left( S_z^2 - \frac{2}{3} \right) + \frac{E}{2} \left( S_+^2 + S_-^2 \right) \text{,}
\end{equation}
where $g_e$ is the electron g-factor, $\mu_e$ is the Bohr magneton, $B_z$ is the external magnetic field parallel to the quantization axis of the central divacancy defect set by the V$_\text{C}$-V$_\text{Si}$ defect axis, $S_z$ is the electron spin $z$ operator, $S_+$ and $S_-$ are the electron spin ladder opearators, and $D$ and $E$ are the zero-field splitting parameters. Due symmetry constraints the $E$ splitting is zero for the C$_{3v}$ symmetric $hh$ and $kk$ configurations, while it is non-zero for the C$_{1h}$ symmetric $kh$ and $hk$ configurations. The experimental zero-field-splitting parameters used in the simulations are reported in Ref.~[\cite{Falk2013}]. The bath Hamiltonian and the coupling Hamiltonian depend on the spin defect considered. For nuclear spins ($I=1/2$) the bath Hamiltonian includes the nuclear Zeeman term only
\begin{equation}
H_{\text{bath}}^{\text{nuc}} = - \sum_i  g_{\text{n}, i} \mu_{\text{N}} B_z I_{z,i} \text{,}
\end{equation}
where $\mu_{\text{N}} $ is the nuclear magneton, $g_{\text{n}, i}$  is the nuclear g-factor of nucleus $i$ being either a $^{29}$Si nucleus or a $^{13}$C nucleus, and $I_{z,i}$ is the nuclear spin $z$ operator of nucleus $i$. For a nuclear spin bath, the coupling Hamiltonian includes the hyperfine term
\begin{equation}
H_{\text{coup}}^{\text{nuc}} = \sum_i S A_i I_i \text{,}
\end{equation}
where the hyperfine tensors $A_i$ are obtained from \emph{ab initio} supercell calculations combined with real space integration methods\cite{IvadyPRb2020} in order to eliminate finite size effects.

The bath Hamiltonian for spin-1/2 point defects can be written as 
\begin{equation}
H_{\text{bath}}^{\text{spin-1/2}} = \sum_i   g_e  \mu_{\text{B}} B_z  \mathcal{S}_{z,i} +  \mathcal{S}_i \mathcal{A}_i \mathcal{I}_i \text{,}
\end{equation}
where the first term is the Zeeman energy of the spin-1/2 bath and the second term is the leading hyperfine interaction for a considered paramagnetic defect. Here, $\mathcal{S}$ and $\mathcal{I}$ are the electron spin operator vector of the environmental defect spin and the nuclear spin operator vector of a nucleus strongly coupled to the defect's electron spin through the hyperfine tensor $\mathcal{A}$, respectively. Based on the discussions in introduction, we consider only the nuclear spin of the $^{14}$N of the shallow nitrogen donor in our study. 

The bath Hamiltonian of environmental divacancy spins contains the Zeeman term and the zero-field-splitting terms as specified in Eq.~(\ref{eq:div}). It is worth mentioning that the quantization axis of the environmental divacancies may not be parallel to the quantization axis of the central divacancy and the external magnetic field. In the random spin bath configurations, we consider all possible arrangements.

According to Ref.~[\cite{Simin2016}] the bath spin Hamiltonian for the quartet silicon vacancy can be written as
\begin{eqnarray} 
H_{\text{bath}}^{\text{spin-3/2}} =\sum_j  \left( D_{\text{VS},j} \left( \mathcal{S}^2_{zj} -\frac{5}{4} \right) + \left[ g_e \mathcal{S}_{zj} + g_{3\parallel} \frac{\mathcal{S}_{+j}^{3} - \mathcal{S}_{-j}^{3}}{4i} \right] \mu_{\text{B}} B_\parallel \right) + \nonumber \\   \sum_{j}  \left( g_e \mu_{\text{B}} \mathcal{S}_{\perp} B_{\perp} +  g_{3\perp} \mu_{\text{B}} \frac{ \left\lbrace \mathcal{S}^2_{+j}, \mathcal{S}_{zj} \right\rbrace B_+ - \left\lbrace \mathcal{S}_{-j}^2 , \mathcal{S}_{zj}^2 \right\rbrace B_- }{2i} \right)  \label{eq:VSIZ}
\end{eqnarray}
where the higher order g-factors $g_{3\parallel}$ and $g_{3\perp}$ are measured in Ref.~[\cite{Simin2016}], the zero-field-splitting constant $D_{\text{VS},j}$ of the environmental silicon vacancy centers is set either to 2.6~MHz or to 35~MHz for the two possible silicon vacancy configurations\cite{IvadyVSi-4H}, $B_\parallel$ and $B_\perp$ are the component of the magnetic field parallel and perpendicular to the C$_{3v}$ axis of the silicon vacancy, and $B_\pm = B_{\perp x} \pm i B_{\perp y}$. For C$_{1h}$ symmetric divacancy configurations, the high symmetry axis and the quantization axis of the silicon vacancy is 109.5~degree aligned to the quantization axis of the central defect and the external magnetic field. 

For point defect spin environments, the coupling Hamiltonian includes dipole-dipole interaction term: 
\begin{equation}
H_{\text{coup}}^{\text{point}} = \sum_i -\frac{\mu_0}{4 \pi} \frac{g_e^2 \mu_{B}^{2}}{ \left| r_{0i} \right|^3} \left( 3 \left(S r_{0i} \right) \left( \mathcal{S}_i r_{0i} \right) - S \mathcal{S}_i \right) \text{,}
\end{equation}
where $r_{0i}$ is the vector pointing from the position of the central divacancy to the position of spin defect $i$. 

We utilize our methodology to identify magnetic field values where spin bath couplings and derived relaxation processes may be resonantly enhanced due to level crossings and cross relaxation effects. In order to capture all resonances, we simulate  high concentrations spin environments, that is the concentration is set to 10$^{18}$~cm$^{-3}$ corresponding to doped or highly irradiated samples. In our quantitative study we focus on the depopulation of the highly populated $m_S = 0$ state due to spin precession and spin flip-flops induced by a spin bath. As the luminescence of divacancies is sensitive to the population of the $m_S = 0$ state,  our results are comparable with magnetic field dependent PL plots.  The intensity of the divacancy luminescence can be estimated from our results  according to the following expression
\begin{equation}
\mathcal{L} \approx  \mathcal{L}_0 \left(  1 -  C p_{m_S=\pm1} \right)  \text{,}
\end{equation}
where $p_{m_S=\pm1} = \varrho_{+1} +  \varrho_{-1}$ is the population of the $m_S = \pm 1$ states,  $C$ is the optical spin contrast, and $ \mathcal{L}_0$ is the luminescence of the bright $m_S=0$ state. 

Furthermore,  we study spin relaxation quantitatively, i.e.\ calculate the spin relaxation rate $T_1^{-1}$, for the most relevant cases. To simulate spin relaxation we do not include additional jump and decoherence operators that would require adjustable rate parameters in the Lindblaian formalism.

\section*{Data and code availability statement}

The data that support the findings of this study and the software package used to carry out the numerical simulations are available from the corresponding author upon reasonable request.

\section*{Acknowledgments}

We acknowledge support from the Knut and Alice Wallenberg Foundation through WBSQD2 project (Grant No.\ 2018.0071). Support from the Swedish Government Strategic Research Area SeRC and the Swedish Government Strategic Research Area in Materials Science on Functional Materials at Linköping University (Faculty Grant SFO-Mat-LiU No. 2009 00971) is gratefully acknowledged. VI acknowledges the support from the MTA Premium Postdoctoral Research Program, the Hungarian NKFIH grants No.\ KKP129866 of the National Excellence Program of Quantum-coherent materials project, and the NKFIH through the National Quantum Technology Program (Grant No. 2017-1.2.1-NKP-2017-00001) and the Quantum Information National Laboratory sponsored by Ministry of Innovation and Technology of Hungary. N. T. S. acknowledges the support from the Swedish Research Council (Grant No. VR 2016-04068), the EU H2020 project QuanTELCO (Grant No. 862721). The calculations were performed on resources provided by the Swedish National Infrastructure for Computing (SNIC) at the National Supercomputer Centre (NSC) partially funded by the Swedish Research Council through grant agreement No. 2018-05973.

\section*{Author contributions}

V.I. and O.B.L. wrote the manuscript with input from all authors. Calculations were performed by O.B.L and V.I. with input from N.T.S. and I.A.A. The computational results were analyzed with contributions from all authors. The research was initiated and supervised by V.I.

\section*{Competing interests}

The authors declare no competing interests.


\begin{thebibliography}{10}
\expandafter\ifx\csname url\endcsname\relax
  \def\url#1{\texttt{#1}}\fi
\expandafter\ifx\csname urlprefix\endcsname\relax\def\urlprefix{URL }\fi
\providecommand{\bibinfo}[2]{#2}
\providecommand{\eprint}[2][]{\url{#2}}

\bibitem{duPreez:1965}
\bibinfo{author}{du~Preez, L.}
\newblock Ph.D. thesis, \bibinfo{school}{University of Witwatersrand}
  (\bibinfo{year}{1965}).

\bibitem{Wrachtrup:JPCM2006}
\bibinfo{author}{Wrachtrup, J.} \& \bibinfo{author}{Jelezko, F.}
\newblock \bibinfo{title}{Processing quantum information in diamond}.
\newblock \emph{\bibinfo{journal}{Journal of Physics-Condensed Matter}}
  \textbf{\bibinfo{volume}{18}}, \bibinfo{pages}{S807--S824}
  (\bibinfo{year}{2006}).

\bibitem{Maze2011}
\bibinfo{author}{Maze, J.~R.} \emph{et~al.}
\newblock \bibinfo{title}{Properties of nitrogen-vacancy centers in diamond:
  the group theoretic approach}.
\newblock \emph{\bibinfo{journal}{New Journal of Physics}}
  \textbf{\bibinfo{volume}{13}}, \bibinfo{pages}{025025}
  (\bibinfo{year}{2011}).

\bibitem{DohertyNVreview}
\bibinfo{author}{Doherty, M.~W.} \emph{et~al.}
\newblock \bibinfo{title}{The nitrogen-vacancy colour centre in diamond}.
\newblock \emph{\bibinfo{journal}{Physics Reports}}
  \textbf{\bibinfo{volume}{528}}, \bibinfo{pages}{1 -- 45}
  (\bibinfo{year}{2013}).


\bibitem{Balasubramanian:Nature2008}
\bibinfo{author}{Balasubramanian, G.} \emph{et~al.}
\newblock \bibinfo{title}{Nanoscale imaging magnetometry with diamond spins
  under ambient conditions}.
\newblock \emph{\bibinfo{journal}{Nature}} \textbf{\bibinfo{volume}{455}},
  \bibinfo{pages}{648--651} (\bibinfo{year}{2008}).

\bibitem{Taylor:NatPhys2008}
\bibinfo{author}{Taylor, J.} \emph{et~al.}
\newblock \bibinfo{title}{High-sensitivity diamond magnetometer with nanoscale
  resolution}.
\newblock \emph{\bibinfo{journal}{Nat. Phys.}} \textbf{\bibinfo{volume}{4}},
  \bibinfo{pages}{810--816} (\bibinfo{year}{2008}).

\bibitem{Kucsko2013}
\bibinfo{author}{Kucsko, G.} \emph{et~al.}
\newblock \bibinfo{title}{Nanometre-scale thermometry in a living cell}.
\newblock \emph{\bibinfo{journal}{Nature}} \textbf{\bibinfo{volume}{500}},
  \bibinfo{pages}{54--58} (\bibinfo{year}{2013}).

\bibitem{PlakhotnikNanoLett2014}
\bibinfo{author}{Plakhotnik, T.}, \bibinfo{author}{Doherty, M.~W.},
  \bibinfo{author}{Cole, J.~H.}, \bibinfo{author}{Chapman, R.} \&
  \bibinfo{author}{Manson, N.~B.}
\newblock \bibinfo{title}{All-{Optical} {Thermometry} and {Thermal}
  {Properties} of the {Optically} {Detected} {Spin} {Resonances} of the {NV}–
  {Center} in {Nanodiamond}}.
\newblock \emph{\bibinfo{journal}{Nano Letters}} \textbf{\bibinfo{volume}{14}},
  \bibinfo{pages}{4989--4996} (\bibinfo{year}{2014}).

\bibitem{Schirhagl2014}
\bibinfo{author}{Schirhagl, R.}, \bibinfo{author}{Chang, K.},
  \bibinfo{author}{Loretz, M.} \& \bibinfo{author}{Degen, C.~L.}
\newblock \bibinfo{title}{Nitrogen-vacancy centers in diamond: Nanoscale
  sensors for physics and biology}.
\newblock \emph{\bibinfo{journal}{Annual Review of Physical Chemistry}}
  \textbf{\bibinfo{volume}{65}}, \bibinfo{pages}{83--105}
  (\bibinfo{year}{2014}).

\bibitem{DegenRMP2017}
\bibinfo{author}{Degen, C.~L.}, \bibinfo{author}{Reinhard, F.} \&
  \bibinfo{author}{Cappellaro, P.}
\newblock \bibinfo{title}{Quantum sensing}.
\newblock \emph{\bibinfo{journal}{Rev. Mod. Phys.}}
  \textbf{\bibinfo{volume}{89}}, \bibinfo{pages}{035002}
  (\bibinfo{year}{2017}).

\bibitem{glenn_high-resolution_2018}
\bibinfo{author}{Glenn, D.~R.} \emph{et~al.}
\newblock \bibinfo{title}{High-resolution magnetic resonance spectroscopy using
  a solid-state spin sensor}.
\newblock \emph{\bibinfo{journal}{Nature}} \textbf{\bibinfo{volume}{555}},
  \bibinfo{pages}{351--354} (\bibinfo{year}{2018}).

\bibitem{schmitt_submillihertz_2017}
\bibinfo{author}{Schmitt, S.} \emph{et~al.}
\newblock \bibinfo{title}{Submillihertz magnetic spectroscopy performed with a
  nanoscale quantum sensor}.
\newblock \emph{\bibinfo{journal}{Science}} \textbf{\bibinfo{volume}{356}},
  \bibinfo{pages}{832--837} (\bibinfo{year}{2017}).

\bibitem{bernien_heralded_2013}
\bibinfo{author}{Bernien, H.} \emph{et~al.}
\newblock \bibinfo{title}{Heralded entanglement between solid-state qubits
  separated by three metres}.
\newblock \emph{\bibinfo{journal}{Nature}} \textbf{\bibinfo{volume}{497}},
  \bibinfo{pages}{86--90} (\bibinfo{year}{2013}).

\bibitem{pfaff_unconditional_2014}
\bibinfo{author}{Pfaff, W.} \emph{et~al.}
\newblock \bibinfo{title}{Unconditional quantum teleportation between distant
  solid-state quantum bits}.
\newblock \emph{\bibinfo{journal}{Science}} \textbf{\bibinfo{volume}{345}},
  \bibinfo{pages}{532--535} (\bibinfo{year}{2014}).


\bibitem{wehner_quantum_2018}
\bibinfo{author}{Wehner, S.}, \bibinfo{author}{Elkouss, D.} \&
  \bibinfo{author}{Hanson, R.}
\newblock \bibinfo{title}{Quantum internet: {A} vision for the road ahead}.
\newblock \emph{\bibinfo{journal}{Science}} \textbf{\bibinfo{volume}{362}}
  (\bibinfo{year}{2018}).

\bibitem{Weber10}
\bibinfo{author}{Weber, J.~R.} \emph{et~al.}
\newblock \bibinfo{title}{Quantum computing with defects}.
\newblock \emph{\bibinfo{journal}{PNAS}} \textbf{\bibinfo{volume}{107}},
  \bibinfo{pages}{8513--8518} (\bibinfo{year}{2010}).


\bibitem{Kurizki2015}
\bibinfo{author}{Kurizki, G.} \emph{et~al.}
\newblock \bibinfo{title}{Quantum technologies with hybrid systems}.
\newblock \emph{\bibinfo{journal}{Proceedings of the National Academy of
  Sciences}} \textbf{\bibinfo{volume}{112}}, \bibinfo{pages}{3866--3873}
  (\bibinfo{year}{2015}).


\bibitem{ZhangPRL2020}
\bibinfo{author}{Zhang, J.}, \bibinfo{author}{Hegde, S.~S.} \&
  \bibinfo{author}{Suter, D.}
\newblock \bibinfo{title}{Efficient implementation of a quantum algorithm in a
  single nitrogen-vacancy center of diamond}.
\newblock \emph{\bibinfo{journal}{Phys. Rev. Lett.}}
  \textbf{\bibinfo{volume}{125}}, \bibinfo{pages}{030501}
  (\bibinfo{year}{2020}).

\bibitem{Koehl11}
\bibinfo{author}{Koehl, W.~F.}, \bibinfo{author}{Buckley, B.~B.},
  \bibinfo{author}{Heremans, F.~J.}, \bibinfo{author}{Calusine, G.} \&
  \bibinfo{author}{Awschalom, D.~D.}
\newblock \bibinfo{title}{Room temperature coherent control of defect spin
  qubits in silicon carbide}.
\newblock \emph{\bibinfo{journal}{Nature}} \textbf{\bibinfo{volume}{479}},
  \bibinfo{pages}{84} (\bibinfo{year}{2011}).

\bibitem{Christle2014}
\bibinfo{author}{Christle, D.~J.} \emph{et~al.}
\newblock \bibinfo{title}{Isolated electron spins in silicon carbide with
  millisecond coherence times}.
\newblock \emph{\bibinfo{journal}{Nat. Mater.}} \textbf{\bibinfo{volume}{14}},
  \bibinfo{pages}{160--163} (\bibinfo{year}{2015}).


\bibitem{Son2006}
\bibinfo{author}{Son, N.~T.} \emph{et~al.}
\newblock \bibinfo{title}{Divacancy in 4h-sic}.
\newblock \emph{\bibinfo{journal}{Phys. Rev. Lett.}}
  \textbf{\bibinfo{volume}{96}}, \bibinfo{pages}{055501}
  (\bibinfo{year}{2006}).


\bibitem{Soltamov12}
\bibinfo{author}{Soltamov, V.~A.}, \bibinfo{author}{Soltamova, A.~A.},
  \bibinfo{author}{Baranov, P.~G.} \& \bibinfo{author}{Proskuryakov, I.~I.}
\newblock \bibinfo{title}{Room temperature coherent spin alignment of silicon
  vacancies in 4h- and 6h-sic}.
\newblock \emph{\bibinfo{journal}{Phys. Rev. Lett.}}
  \textbf{\bibinfo{volume}{108}}, \bibinfo{pages}{226402}
  (\bibinfo{year}{2012}).

\bibitem{Widmann2014}
\bibinfo{author}{Widmann, M.} \emph{et~al.}
\newblock \bibinfo{title}{Coherent control of single spins in silicon carbide
  at room temperature}.
\newblock \emph{\bibinfo{journal}{Nat. Mater.}} \textbf{\bibinfo{volume}{14}},
  \bibinfo{pages}{164--168} (\bibinfo{year}{2015}).

\bibitem{miao_universal_2020}
\bibinfo{author}{Miao, K.~C.} \emph{et~al.}
\newblock \bibinfo{title}{Universal coherence protection in a solid-state spin
  qubit}.
\newblock \emph{\bibinfo{journal}{Science}} \textbf{\bibinfo{volume}{369}},
  \bibinfo{pages}{1493--1497} (\bibinfo{year}{2020}).


\bibitem{ChristlePRX2017}
\bibinfo{author}{Christle, D.~J.} \emph{et~al.}
\newblock \bibinfo{title}{Isolated spin qubits in sic with a high-fidelity
  infrared spin-to-photon interface}.
\newblock \emph{\bibinfo{journal}{Phys. Rev. X}} \textbf{\bibinfo{volume}{7}},
  \bibinfo{pages}{021046} (\bibinfo{year}{2017}).


\bibitem{FalkPRL2015}
\bibinfo{author}{Falk, A.~L.} \emph{et~al.}
\newblock \bibinfo{title}{Optical polarization of nuclear spins in silicon
  carbide}.
\newblock \emph{\bibinfo{journal}{Phys. Rev. Lett.}}
  \textbf{\bibinfo{volume}{114}}, \bibinfo{pages}{247603}
  (\bibinfo{year}{2015}).

\bibitem{bourassa_entanglement_2020}
\bibinfo{author}{Bourassa, A.} \emph{et~al.}
\newblock \bibinfo{title}{Entanglement and control of single nuclear spins in
  isotopically engineered silicon carbide}.
\newblock \emph{\bibinfo{journal}{Nature Materials}}
  \textbf{\bibinfo{volume}{19}}, \bibinfo{pages}{1319--1325}
  (\bibinfo{year}{2020}).


\bibitem{IvadyDNP2015}
\bibinfo{author}{Iv\'ady, V.} \emph{et~al.}
\newblock \bibinfo{title}{Theoretical model of dynamic spin polarization of
  nuclei coupled to paramagnetic point defects in diamond and silicon carbide}.
\newblock \emph{\bibinfo{journal}{Phys. Rev. B}} \textbf{\bibinfo{volume}{92}},
  \bibinfo{pages}{115206} (\bibinfo{year}{2015}).


\bibitem{IvadyPRL2016}
\bibinfo{author}{Iv\'ady, V.} \emph{et~al.}
\newblock \bibinfo{title}{High-fidelity bidirectional nuclear qubit
  initialization in sic}.
\newblock \emph{\bibinfo{journal}{Phys. Rev. Lett.}}
  \textbf{\bibinfo{volume}{117}}, \bibinfo{pages}{220503}
  (\bibinfo{year}{2016}).

\bibitem{li_room_2020}
\bibinfo{author}{Li, Q.} \emph{et~al.}
\newblock \bibinfo{title}{Room temperature coherent manipulation of single-spin
  qubits in silicon carbide with high readout contrast}.
\newblock \emph{\bibinfo{journal}{arXiv:2005.07876}}  (\bibinfo{year}{2020}).


\bibitem{whiteley_spinphonon_2019}
\bibinfo{author}{Whiteley, S.~J.} \emph{et~al.}
\newblock \bibinfo{title}{Spin–phonon interactions in silicon carbide
  addressed by {Gaussian} acoustics}.
\newblock \emph{\bibinfo{journal}{Nature Physics}}
  \textbf{\bibinfo{volume}{15}}, \bibinfo{pages}{490--495}
  (\bibinfo{year}{2019}).

\bibitem{anderson_electrical_2019}
\bibinfo{author}{Anderson, C.~P.} \emph{et~al.}
\newblock \bibinfo{title}{Electrical and optical control of single spins
  integrated in scalable semiconductor devices}.
\newblock \emph{\bibinfo{journal}{Science}} \textbf{\bibinfo{volume}{366}},
  \bibinfo{pages}{1225--1230} (\bibinfo{year}{2019}).


\bibitem{son_developing_2020}
\bibinfo{author}{Son, N.~T.} \emph{et~al.}
\newblock \bibinfo{title}{Developing silicon carbide for quantum spintronics}.
\newblock \emph{\bibinfo{journal}{Applied Physics Letters}}
  \textbf{\bibinfo{volume}{116}}, \bibinfo{pages}{190501}
  (\bibinfo{year}{2020}).

\bibitem{pham_enhanced_2012}
\bibinfo{author}{Pham, L.~M.} \emph{et~al.}
\newblock \bibinfo{title}{Enhanced solid-state multispin metrology using
  dynamical decoupling}.
\newblock \emph{\bibinfo{journal}{Physical Review B}}
  \textbf{\bibinfo{volume}{86}}, \bibinfo{pages}{045214}
  (\bibinfo{year}{2012}).

\bibitem{bar-gill_solid-state_2013}
\bibinfo{author}{Bar-Gill, N.}, \bibinfo{author}{Pham, L.~M.},
  \bibinfo{author}{Jarmola, A.}, \bibinfo{author}{Budker, D.} \&
  \bibinfo{author}{Walsworth, R.~L.}
\newblock \bibinfo{title}{Solid-state electronic spin coherence time
  approaching one second}.
\newblock \emph{\bibinfo{journal}{Nature Communications}}
  \textbf{\bibinfo{volume}{4}}, \bibinfo{pages}{1743} (\bibinfo{year}{2013}).


\bibitem{RomachPRL2015}
\bibinfo{author}{Romach, Y.} \emph{et~al.}
\newblock \bibinfo{title}{Spectroscopy of surface-induced noise using shallow
  spins in diamond}.
\newblock \emph{\bibinfo{journal}{Phys. Rev. Lett.}}
  \textbf{\bibinfo{volume}{114}}, \bibinfo{pages}{017601}
  (\bibinfo{year}{2015}).


\bibitem{barry_sensitivity_2020}
\bibinfo{author}{Barry, J.~F.} \emph{et~al.}
\newblock \bibinfo{title}{Sensitivity optimization for {NV}-diamond
  magnetometry}.
\newblock \emph{\bibinfo{journal}{Reviews of Modern Physics}}
  \textbf{\bibinfo{volume}{92}}, \bibinfo{pages}{015004}
  (\bibinfo{year}{2020}).


\bibitem{takahashi_quenching_2008}
\bibinfo{author}{Takahashi, S.}, \bibinfo{author}{Hanson, R.},
  \bibinfo{author}{van Tol, J.}, \bibinfo{author}{Sherwin, M.~S.} \&
  \bibinfo{author}{Awschalom, D.~D.}
\newblock \bibinfo{title}{Quenching {Spin} {Decoherence} in {Diamond} through
  {Spin} {Bath} {Polarization}}.
\newblock \emph{\bibinfo{journal}{Physical Review Letters}}
  \textbf{\bibinfo{volume}{101}}, \bibinfo{pages}{047601}
  (\bibinfo{year}{2008}).


\bibitem{jarmola_temperature_2012}
\bibinfo{author}{Jarmola, A.}, \bibinfo{author}{Acosta, V.~M.},
  \bibinfo{author}{Jensen, K.}, \bibinfo{author}{Chemerisov, S.} \&
  \bibinfo{author}{Budker, D.}
\newblock \bibinfo{title}{Temperature- and {Magnetic}-{Field}-{Dependent}
  {Longitudinal} {Spin} {Relaxation} in {Nitrogen}-{Vacancy} {Ensembles} in
  {Diamond}}.
\newblock \emph{\bibinfo{journal}{Physical Review Letters}}
  \textbf{\bibinfo{volume}{108}}, \bibinfo{pages}{197601}
  (\bibinfo{year}{2012}).


\bibitem{Wang2014}
\bibinfo{author}{Wang, H.-J.} \emph{et~al.}
\newblock \bibinfo{title}{Optically detected cross-relaxation spectroscopy of
  electron spins in diamond}.
\newblock \emph{\bibinfo{journal}{Nature Communications}}
  \textbf{\bibinfo{volume}{5}}, \bibinfo{pages}{4135} (\bibinfo{year}{2014}).


\bibitem{WickenbrockAPL2016}
\bibinfo{author}{Wickenbrock, A.} \emph{et~al.}
\newblock \bibinfo{title}{Microwave-free magnetometry with nitrogen-vacancy
  centers in diamond}.
\newblock \emph{\bibinfo{journal}{Applied Physics Letters}}
  \textbf{\bibinfo{volume}{109}}, \bibinfo{pages}{053505}
  (\bibinfo{year}{2016}).


\bibitem{WoodPRB2016}
\bibinfo{author}{Wood, J. D.~A.} \emph{et~al.}
\newblock \bibinfo{title}{Wide-band nanoscale magnetic resonance spectroscopy
  using quantum relaxation of a single spin in diamond}.
\newblock \emph{\bibinfo{journal}{Phys. Rev. B}} \textbf{\bibinfo{volume}{94}},
  \bibinfo{pages}{155402} (\bibinfo{year}{2016}).


\bibitem{WoodNatComm2017}
\bibinfo{author}{Wood, J. D.~A.} \emph{et~al.}
\newblock \bibinfo{title}{Microwave-free nuclear magnetic resonance at
  molecular scales}.
\newblock \emph{\bibinfo{journal}{Nature Communications}}
  \textbf{\bibinfo{volume}{8}}, \bibinfo{pages}{15950} (\bibinfo{year}{2017}).
  

\bibitem{ivady_photoluminescence_2021}
\bibinfo{author}{Ivády, V.} \emph{et~al.}
\newblock \bibinfo{title}{Photoluminescence at the ground-state level
  anticrossing of the nitrogen-vacancy center in diamond: {A} comprehensive
  study}.
\newblock \emph{\bibinfo{journal}{Physical Review B}}
  \textbf{\bibinfo{volume}{103}}, \bibinfo{pages}{035307}
  (\bibinfo{year}{2021}).


\bibitem{koehl_room_2011}
\bibinfo{author}{Koehl, W.~F.}, \bibinfo{author}{Buckley, B.~B.},
  \bibinfo{author}{Heremans, F.~J.}, \bibinfo{author}{Calusine, G.} \&
  \bibinfo{author}{Awschalom, D.~D.}
\newblock \bibinfo{title}{Room temperature coherent control of defect spin
  qubits in silicon carbide}.
\newblock \emph{\bibinfo{journal}{Nature}} \textbf{\bibinfo{volume}{479}},
  \bibinfo{pages}{84--87} (\bibinfo{year}{2011}).


\bibitem{falk_polytype_2013}
\bibinfo{author}{Falk, A.~L.} \emph{et~al.}
\newblock \bibinfo{title}{Polytype control of spin qubits in silicon carbide}.
\newblock \emph{\bibinfo{journal}{Nature Communications}}
  \textbf{\bibinfo{volume}{4}}, \bibinfo{pages}{1819} (\bibinfo{year}{2013}).


\bibitem{yan_room-temperature_2020}
\bibinfo{author}{Yan, F.-F.} \emph{et~al.}
\newblock \bibinfo{title}{Room-temperature coherent control of implanted defect
  spins in silicon carbide}.
\newblock \emph{\bibinfo{journal}{npj Quantum Information}}
  \textbf{\bibinfo{volume}{6}}, \bibinfo{pages}{1--6} (\bibinfo{year}{2020}).


\bibitem{yang_quantum_2008}
\bibinfo{author}{Yang, W.} \& \bibinfo{author}{Liu, R.-B.}
\newblock \bibinfo{title}{Quantum many-body theory of qubit decoherence in a
  finite-size spin bath}.
\newblock \emph{\bibinfo{journal}{Physical Review B}}
  \textbf{\bibinfo{volume}{78}}, \bibinfo{pages}{085315}
  (\bibinfo{year}{2008}).


\bibitem{seo_quantum_2016}
\bibinfo{author}{Seo, H.} \emph{et~al.}
\newblock \bibinfo{title}{Quantum decoherence dynamics of divacancy spins in
  silicon carbide}.
\newblock \emph{\bibinfo{journal}{Nature Communications}}
  \textbf{\bibinfo{volume}{7}}, \bibinfo{pages}{12935} (\bibinfo{year}{2016}).


\bibitem{onizhuk_probing_2021}
\bibinfo{author}{Onizhuk, M.} \emph{et~al.}
\newblock \bibinfo{title}{Probing the {Coherence} of {Solid}-{State} {Qubits}
  at {Avoided} {Crossings}}.
\newblock \emph{\bibinfo{journal}{PRX Quantum}} \textbf{\bibinfo{volume}{2}},
  \bibinfo{pages}{010311} (\bibinfo{year}{2021}).


\bibitem{shrivastava_theory_1983}
\bibinfo{author}{Shrivastava, K.~N.}
\newblock \bibinfo{title}{Theory of {Spin}–{Lattice} {Relaxation}}.
\newblock \emph{\bibinfo{journal}{physica status solidi (b)}}
  \textbf{\bibinfo{volume}{117}}, \bibinfo{pages}{437--458}
  (\bibinfo{year}{1983}).


\bibitem{norambuena_spin-lattice_2018}
\bibinfo{author}{Norambuena, A.} \emph{et~al.}
\newblock \bibinfo{title}{Spin-lattice relaxation of individual solid-state
  spins}.
\newblock \emph{\bibinfo{journal}{Physical Review B}}
  \textbf{\bibinfo{volume}{97}}, \bibinfo{pages}{094304}
  (\bibinfo{year}{2018}).


\bibitem{radczyk_applications_2003}
\bibinfo{author}{Radczyk, T.}, \bibinfo{author}{Hoffmann, S.} \&
  \bibinfo{author}{Goslar, J.}
\newblock \bibinfo{title}{Applications of the {Transport} {Integrals} in
  {Solid}-{State} {Physics} and in {Electron} {Spin} {Relaxation}}.
\newblock \emph{\bibinfo{journal}{Acta Physica Polonica A}}
  \textbf{\bibinfo{volume}{5}}, \bibinfo{pages}{469--477}
  (\bibinfo{year}{2003}).


\bibitem{hoffmann_raman_2013}
\bibinfo{author}{Hoffmann, S.~K.} \& \bibinfo{author}{Lijewski, S.}
\newblock \bibinfo{title}{Raman electron spin–lattice relaxation with the
  {Debye}-type and with real phonon spectra in crystals}.
\newblock \emph{\bibinfo{journal}{Journal of Magnetic Resonance}}
  \textbf{\bibinfo{volume}{227}}, \bibinfo{pages}{51--56}
  (\bibinfo{year}{2013}).


\bibitem{redman_spin_1991}
\bibinfo{author}{Redman, D.~A.}, \bibinfo{author}{Brown, S.},
  \bibinfo{author}{Sands, R.~H.} \& \bibinfo{author}{Rand, S.~C.}
\newblock \bibinfo{title}{Spin dynamics and electronic states of {N}-{V}
  centers in diamond by {EPR} and four-wave-mixing spectroscopy}.
\newblock \emph{\bibinfo{journal}{Physical Review Letters}}
  \textbf{\bibinfo{volume}{67}}, \bibinfo{pages}{3420--3423}
  (\bibinfo{year}{1991}).


\bibitem{simin_locking_2017}
\bibinfo{author}{Simin, D.} \emph{et~al.}
\newblock \bibinfo{title}{Locking of electron spin coherence above 20 ms in
  natural silicon carbide}.
\newblock \emph{\bibinfo{journal}{Physical Review B}}
  \textbf{\bibinfo{volume}{95}}, \bibinfo{pages}{161201}
  (\bibinfo{year}{2017}).


\bibitem{soltamov_relaxation_2021}
\bibinfo{author}{Soltamov, V.~A.} \emph{et~al.}
\newblock \bibinfo{title}{Relaxation processes and high-field coherent spin
  manipulation in color center ensembles in {6H}-{SiC}}.
\newblock \emph{\bibinfo{journal}{Physical Review B}}
  \textbf{\bibinfo{volume}{103}}, \bibinfo{pages}{195201}
  (\bibinfo{year}{2021}).


\bibitem{lin_temperature_2021}
\bibinfo{author}{Lin, W.-X.} \emph{et~al.}
\newblock \bibinfo{title}{Temperature dependence of divacancy spin coherence in
  implanted silicon carbide}.
\newblock \emph{\bibinfo{journal}{arXiv:2104.12089 [quant-ph]}}
  (\bibinfo{year}{2021}).


\bibitem{restrepo_full_2012}
\bibinfo{author}{Restrepo, O.~D.} \& \bibinfo{author}{Windl, W.}
\newblock \bibinfo{title}{Full {First}-{Principles} {Theory} of {Spin}
  {Relaxation} in {Group}-{IV} {Materials}}.
\newblock \emph{\bibinfo{journal}{Physical Review Letters}}
  \textbf{\bibinfo{volume}{109}}, \bibinfo{pages}{166604}
  (\bibinfo{year}{2012}).


\bibitem{gugler_ab_2018}
\bibinfo{author}{Gugler, J.} \emph{et~al.}
\newblock \bibinfo{title}{\textit{{Ab} initio} calculation of the spin lattice
  relaxation time {T} 1 for nitrogen-vacancy centers in diamond}.
\newblock \emph{\bibinfo{journal}{Physical Review B}}
  \textbf{\bibinfo{volume}{98}}, \bibinfo{pages}{214442}
  (\bibinfo{year}{2018}).


\bibitem{astner_solid-state_2018}
\bibinfo{author}{Astner, T.} \emph{et~al.}
\newblock \bibinfo{title}{Solid-state electron spin lifetime limited by
  phononic vacuum modes}.
\newblock \emph{\bibinfo{journal}{Nature Materials}}
  \textbf{\bibinfo{volume}{17}}, \bibinfo{pages}{313--317}
  (\bibinfo{year}{2018}).


\bibitem{park_spin-phonon_2020}
\bibinfo{author}{Park, J.}, \bibinfo{author}{Zhou, J.-J.} \&
  \bibinfo{author}{Bernardi, M.}
\newblock \bibinfo{title}{Spin-phonon relaxation times in centrosymmetric
  materials from first principles}.
\newblock \emph{\bibinfo{journal}{Physical Review B}}
  \textbf{\bibinfo{volume}{101}}, \bibinfo{pages}{045202}
  (\bibinfo{year}{2020}).
  

\bibitem{xu_spin-phonon_2020}
\bibinfo{author}{Xu, J.} \emph{et~al.}
\newblock \bibinfo{title}{Spin-phonon relaxation from a universal ab initio
  density-matrix approach}.
\newblock \emph{\bibinfo{journal}{Nature Communications}}
  \textbf{\bibinfo{volume}{11}}, \bibinfo{pages}{2780} (\bibinfo{year}{2020}).
  

\bibitem{IvadyPRb2020}
\bibinfo{author}{Iv\'ady, V.}
\newblock \bibinfo{title}{Longitudinal spin relaxation model applied to
  point-defect qubit systems}.
\newblock \emph{\bibinfo{journal}{Phys. Rev. B}}
  \textbf{\bibinfo{volume}{101}}, \bibinfo{pages}{155203}
  (\bibinfo{year}{2020}).


\bibitem{kawanishi_effect_2016}
\bibinfo{author}{Kawanishi, S.} \& \bibinfo{author}{Mizoguchi, T.}
\newblock \bibinfo{title}{Effect of van der {Waals} interactions on the
  stability of {SiC} polytypes}.
\newblock \emph{\bibinfo{journal}{Journal of Applied Physics}}
  \textbf{\bibinfo{volume}{119}}, \bibinfo{pages}{175101}
  (\bibinfo{year}{2016}).


\bibitem{davidsson_first_2018}
\bibinfo{author}{Davidsson, J.} \emph{et~al.}
\newblock \bibinfo{title}{First principles predictions of magneto-optical data
  for semiconductor point defect identification: the case of divacancy defects
  in {4H}–{SiC}}.
\newblock \emph{\bibinfo{journal}{New Journal of Physics}}
  \textbf{\bibinfo{volume}{20}}, \bibinfo{pages}{023035}
  (\bibinfo{year}{2018}).


\bibitem{MagnussonPRB2018}
\bibinfo{author}{Magnusson, B.} \emph{et~al.}
\newblock \bibinfo{title}{Excitation properties of the divacancy in $4h$-sic}.
\newblock \emph{\bibinfo{journal}{Phys. Rev. B}} \textbf{\bibinfo{volume}{98}},
  \bibinfo{pages}{195202} (\bibinfo{year}{2018}).


\bibitem{ivady_stabilization_2019}
\bibinfo{author}{Iv\'{a}dy, V.} \emph{et~al.}
\newblock \bibinfo{title}{Stabilization of point-defect spin qubits by quantum
  wells}.
\newblock \emph{\bibinfo{journal}{Nature Communications}}
  \textbf{\bibinfo{volume}{10}}, \bibinfo{pages}{5607} (\bibinfo{year}{2019}).
  

\bibitem{nagy_high-fidelity_2019}
\bibinfo{author}{Nagy, R.} \emph{et~al.}
\newblock \bibinfo{title}{High-fidelity spin and optical control of single
  silicon-vacancy centres in silicon carbide}.
\newblock \emph{\bibinfo{journal}{Nature Communications}}
  \textbf{\bibinfo{volume}{10}}, \bibinfo{pages}{1954} (\bibinfo{year}{2019}).


\bibitem{kasper_influence_2020}
\bibinfo{author}{Kasper, C.} \emph{et~al.}
\newblock \bibinfo{title}{Influence of {Irradiation} on {Defect} {Spin}
  {Coherence} in {Silicon} {Carbide}}.
\newblock \emph{\bibinfo{journal}{Physical Review Applied}}
  \textbf{\bibinfo{volume}{13}}, \bibinfo{pages}{044054}
  (\bibinfo{year}{2020}).


\bibitem{Son2007}
\bibinfo{author}{Son, N.~T.}, \bibinfo{author}{Carlsson, P.},
  \bibinfo{author}{ul~Hassan, J.}, \bibinfo{author}{Magnusson, B.} \&
  \bibinfo{author}{Janz\'en, E.}
\newblock \bibinfo{title}{Defects and carrier compensation in semi-insulating
  $4h\text{\ensuremath{-}}\mathrm{Si}\mathrm{C}$ substrates}.
\newblock \emph{\bibinfo{journal}{Phys. Rev. B}} \textbf{\bibinfo{volume}{75}},
  \bibinfo{pages}{155204} (\bibinfo{year}{2007}).


\bibitem{greulichweber_epr_1997}
\bibinfo{author}{Greulich-Weber, S.}
\newblock \bibinfo{title}{{EPR} and {ENDOR} {Investigations} of {Shallow}
  {Impurities} in {SiC} {Polytypes}}.
\newblock \emph{\bibinfo{journal}{physica status solidi (a)}}
  \textbf{\bibinfo{volume}{162}}, \bibinfo{pages}{95--151}
  (\bibinfo{year}{1997}).

\bibitem{DefoPRB2018}
\bibinfo{author}{Kuate~Defo, R.} \emph{et~al.}
\newblock \bibinfo{title}{Energetics and kinetics of vacancy defects in
  $4h$-sic}.
\newblock \emph{\bibinfo{journal}{Phys. Rev. B}} \textbf{\bibinfo{volume}{98}},
  \bibinfo{pages}{104103} (\bibinfo{year}{2018}).


\bibitem{IvadyVSi-4H}
\bibinfo{author}{Iv\'ady, V.} \emph{et~al.}
\newblock \bibinfo{title}{Identification of si-vacancy related room-temperature
  qubits in $4h$ silicon carbide}.
\newblock \emph{\bibinfo{journal}{Phys. Rev. B}} \textbf{\bibinfo{volume}{96}},
  \bibinfo{pages}{161114} (\bibinfo{year}{2017}).


\bibitem{Simin2016}
\bibinfo{author}{Simin, D.} \emph{et~al.}
\newblock \bibinfo{title}{All-optical dc nanotesla magnetometry using silicon
  vacancy fine structure in isotopically purified silicon carbide}.
\newblock \emph{\bibinfo{journal}{Phys. Rev. X}} \textbf{\bibinfo{volume}{6}},
  \bibinfo{pages}{031014} (\bibinfo{year}{2016}).


\end{thebibliography}

\end{document}